\pdfoutput=1
\documentclass{article}

\usepackage[preprint]{neurips_data_2023}

\usepackage[utf8]{inputenc} 
\usepackage[T1]{fontenc}    
\PassOptionsToPackage{hyphens}{url}\usepackage{hyperref}
\usepackage{url}            
\usepackage{booktabs}       
\usepackage{amsfonts}       
\usepackage{nicefrac}       
\usepackage{microtype}      
\usepackage{xcolor}         
\usepackage{comment}
\usepackage{graphicx}
\usepackage{longtable}
\usepackage{soul}
\usepackage{tikz}
\usetikzlibrary{positioning, shapes, backgrounds, calc}
\usepackage{lmodern}
\usepackage{caption}
\usepackage{pgfplots}
\usepackage{float}
\usepackage{multirow}
\usepackage{pifont}

\definecolor{graphicbackground}{rgb}{0.96,0.96,0.8}
\definecolor{codebackground}{rgb}{0.9,0.9,1}

\newcommand{\cmark}{\ding{51}}%
\newcommand{\xmark}{\ding{55}}%

\usepackage{array}
\newcolumntype{L}[1]{>{\raggedright\let\newline\\\arraybackslash\hspace{0pt}}m{#1}}
\newcolumntype{C}[1]{>{\centering\let\newline\\\arraybackslash\hspace{0pt}}m{#1}}
\newcolumntype{R}[1]{>{\raggedleft\let\newline\\\arraybackslash\hspace{0pt}}m{#1}}

\newcommand{\obelisk}{\hspace{-0.5cm}{\includegraphics[height=1.13cm]{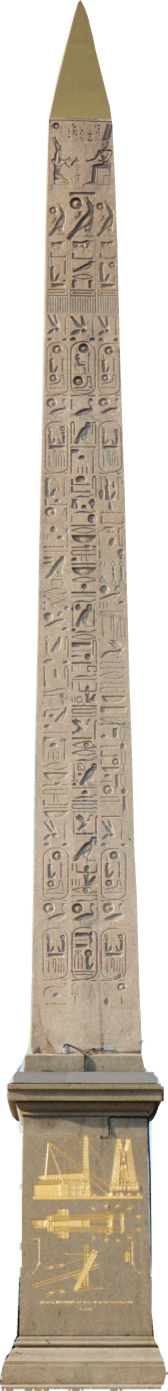}}}

\newcommand{\obelics}{\hspace{-1.5cm}{\includegraphics[height=1.3cm]{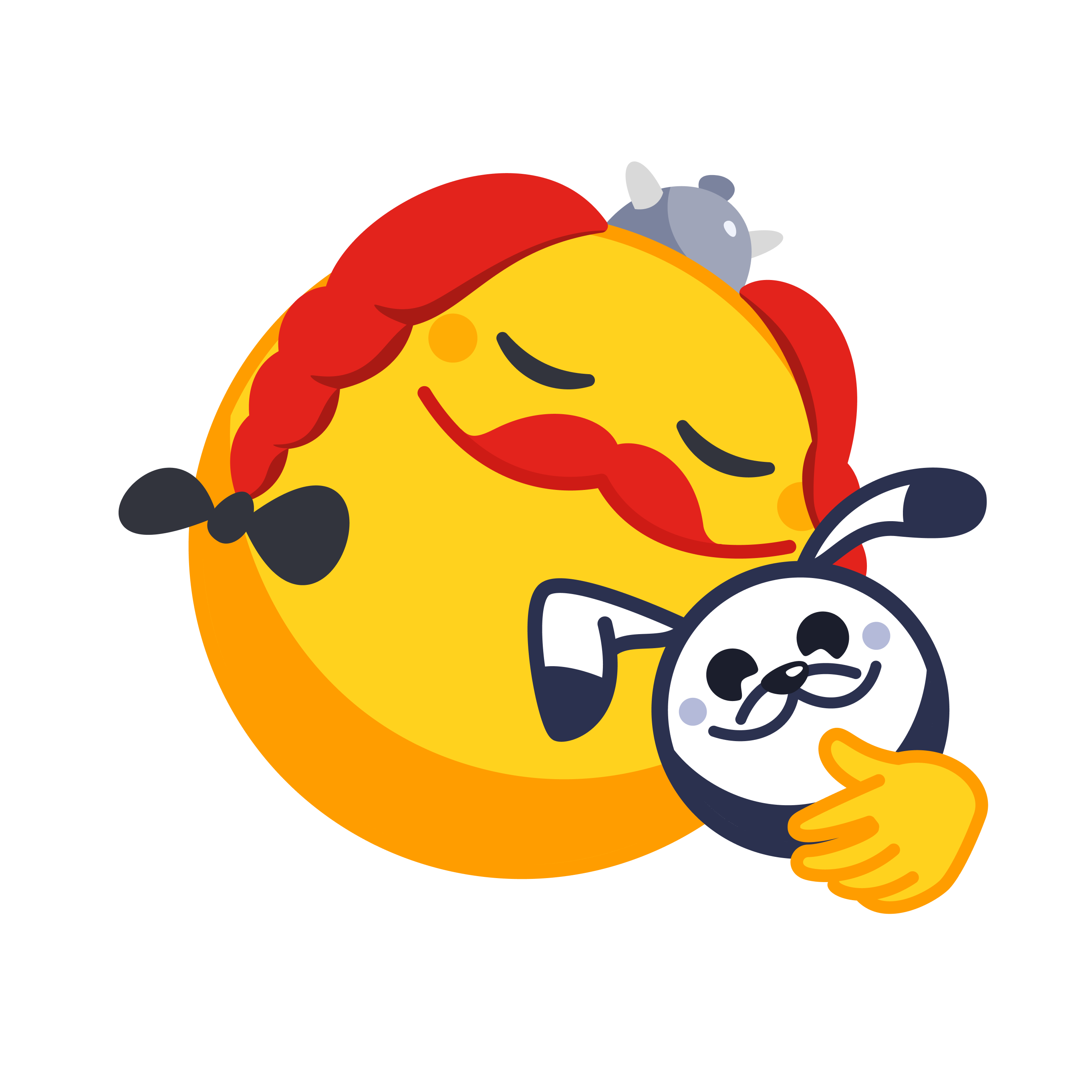}}}

\title{\obelisk OBELICS: An Open Web-Scale Filtered Dataset of Interleaved Image-Text Documents}

\title{
  \begin{minipage}[t]{0.01\linewidth}
    \vspace*{-3pt}
    \obelics
  \end{minipage}%
  \begin{minipage}[t]{0.99\linewidth}
    \vspace*{0pt}
    \centering
    OBELICS: An Open Web-Scale Filtered Dataset of Interleaved Image-Text Documents
  \end{minipage}
}

\author{%
Hugo Laurençon$^{*,1,2}$ \quad Lucile Saulnier$^{*,1}$ \quad Léo Tronchon$^{*,1}$\\
\textbf{Stas Bekman}$^{*,1}$ \quad \textbf{Amanpreet Singh}$^{*,1}$ \quad \textbf{Anton Lozhkov}$^{1}$\\
\textbf{Thomas Wang}$^{1}$ \quad \textbf{Siddharth Karamcheti}$^{1,3}$ \quad \textbf{Alexander M. Rush}$^{\dag,1}$\\
\textbf{Douwe Kiela}$^{\dag,1,3}$ \quad \textbf{Matthieu Cord}$^{\dag,2}$ \quad \textbf{Victor Sanh}$^{*,\dag,1}$\\
$^{*}$Equal contributions, $^{\dag}$Senior contributions\vspace{0.3em}\\
\texttt{hugo@huggingface.co}\vspace{0.3em}\\
$^1$Hugging Face   $^2$Sorbonne Université   $^3$Stanford University
}

\pgfplotsset{compat=1.18}
\begin{document}

\maketitle

\begin{abstract} 
Large multimodal models trained on natural documents, which interleave images and text, outperform models trained on image-text pairs on various multimodal benchmarks. However, the datasets used to train these models have not been released, and the collection process has not been fully specified.  We introduce the \texttt{OBELICS} dataset, an open web-scale filtered dataset of interleaved image-text documents comprising 141 million web pages extracted from Common Crawl, 353 million associated images, and 115 billion text tokens. We describe the dataset creation process, present comprehensive filtering rules, and provide an analysis of the dataset's content. To show the viability of \texttt{OBELICS}, we train vision and language models of 9 and 80 billion parameters named \texttt{IDEFICS}, and obtain competitive performance on different multimodal benchmarks. We release our dataset, models and code.\footnote{\parbox{\linewidth}{\texttt{OBELICS}: \url{https://huggingface.co/datasets/HuggingFaceM4/OBELICS}\\\texttt{OBELICS} reproduction code: \url{https://github.com/huggingface/OBELICS}\\\texttt{IDEFICS} models: \url{https://huggingface.co/HuggingFaceM4/idefics-80b}}}.
\end{abstract}

\begin{figure}[h]
\centering
\includegraphics[width=0.9\textwidth]{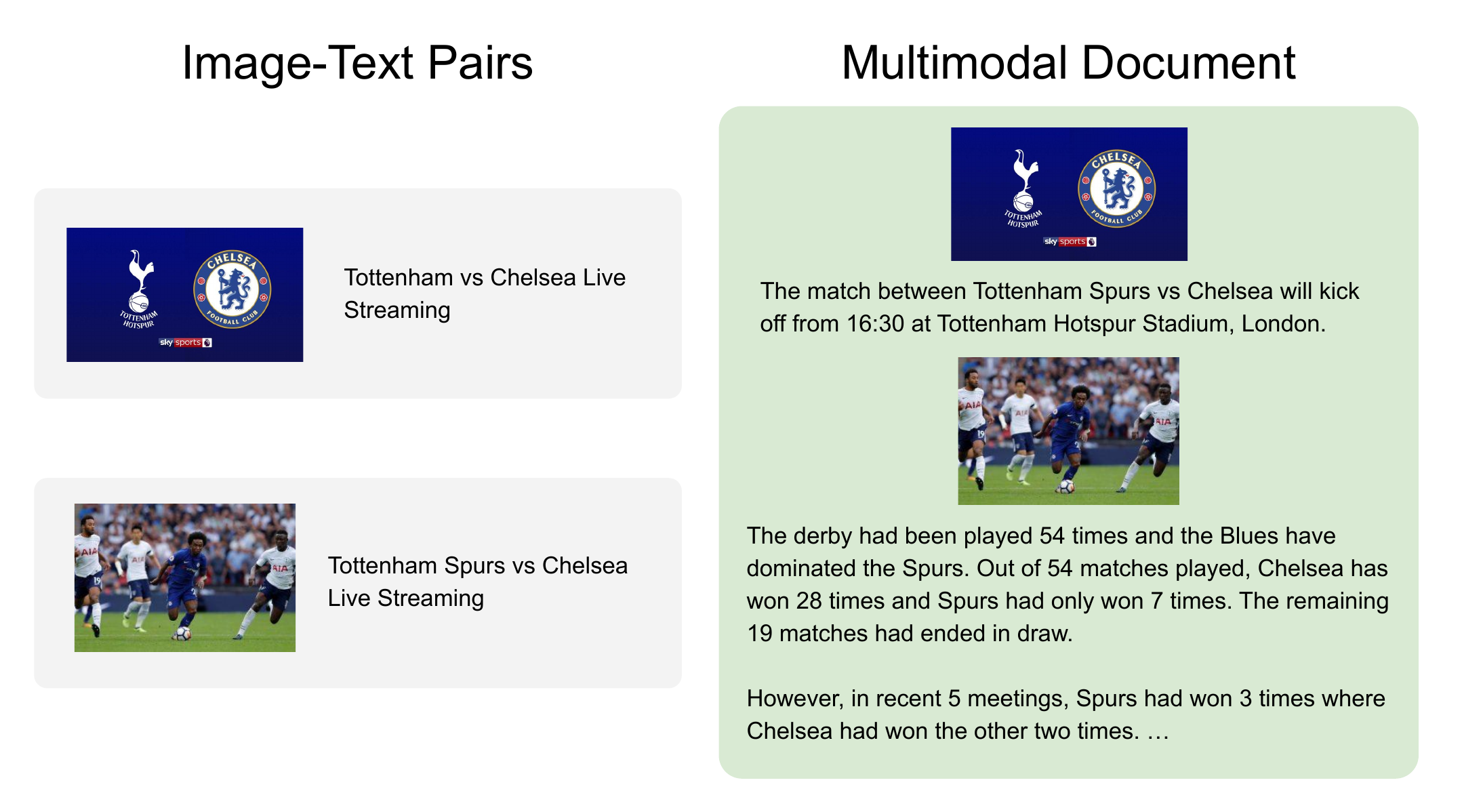}
\caption{A comparison of extraction from the same web document. For image-text pairs, the alt-text of images is often short or non-grammatical. For \texttt{OBELICS}, the extracted multimodal web document interleaves long-form text with the images on the page.}
\label{fig:viz_web_doc}
\end{figure}

\section{Introduction}

Recent systems demonstrate the effectiveness of training large multimodal models such as Flamingo on naturally occurring multimodal documents~\citep{Flamingo,CM3,KOSMOS}. A multimodal document is a succession of text paragraphs interleaved by images, such as web pages that contain images. Models trained on these web documents outperform vision and language models trained solely on image-text pairs on various benchmarks \citep{Flamingo}. They can also generate long and coherent text about a set of multiple images.

While these results are compelling, they have not been replicable. The datasets used in these works are not publicly available,  and relatively little information is known about their creation process and composition. This state motivates the creation of large-scale collections of high-quality multimodal web documents to support the creation of the next generation of models. 

We take inspiration from existing large open image-text datasets such as LAION \citep{LAION-5B} and COYO \citep{COYO}, comprised of hundreds of millions of image-text pairs obtained through web crawling.
These datasets have been critical to developing and replicating numerous recent multimodal models~\citep{CLIP, BEIT3, CoCa, OFA, Prismer}. While this approach allows for building extremely large and diverse training datasets, we note several limitations to using only image-text pairs. From a language perspective, these datasets rely primarily on alt-text, meaning the text given is brief, captures an approximate snapshot of the image's content, and often lacks grammatical correctness. From a document perspective, image-text pairs remove an image from its natural context on a page and its relationship with other documents. 

In this work, we introduce \texttt{OBELICS}\footnote{Open Bimodal Examples from Large fIltered Commoncrawl Snapshots}, an openly-accessible curated web-scale dataset consisting of {141 million multimodal English web documents} which contain {353 million associated images} and {115 billion tokens}. \texttt{OBELICS} collects full multimodal documents interleaving text and images as shown in Figure~\ref{fig:viz_web_doc}. We describe the dataset creation process, outline the filtering and curation steps and shed light on the dataset's content and limitations. To demonstrate the viability of \texttt{OBELICS}, we train \texttt{IDEFICS}, an 80 billion parameter multimodal model and show competitive performance against large-scale multimodal models such as Flamingo \citep{Flamingo}.

\section{Related Works}

\paragraph{Image-text pairs datasets} The largest multimodal datasets, such as LAION \citep{LAION-400M, LAION-5B}, Conceptual Captions \citep{CC3M, CC12M}, ALIGN \citep{ALIGN}, COYO \citep{COYO}, and DataComp \citep{DataComp}, contain billions of image-text pairs and are usually obtained through web-crawling and alt-text extraction. A variety of multimodal models have been trained on this type of dataset: multimodal encoder models which use a contrastive objective \citep{CLIP,BEIT3}, image generation based on Transformers or diffusion processes \citep{GLIDE, DallE2, Latent_diffusion_model, Imagen}. While the scale of these datasets makes them attractive candidates for training, our work focuses on extracting images and the textual context in which they appear instead of extracting the associated alternative text. 

\paragraph{Web document datasets}
Insights from scaling language models \citep{OpenAIScalingLaws, Chinchilla} emphasize the need for increasingly bigger datasets. For instance, LLaMA \citep{LLaMA} was trained on a dataset of 1.4T tokens created exclusively from openly accessible English web content. The authors noticed that an even bigger dataset would have benefited the model. To address that need, multiple web-scale datasets have been introduced and made available: c4 \citep{T5_C4}, ROOTS \citep{ROOTS}, Pile \citep{Pile}, OSCAR \citep{OSCAR}. Although \texttt{OBELICS} falls in the same category of making accessible large collections of curated web documents, the additional extraction of images changes the nature of the resulting dataset. It allows training models with additional vision capabilities.

\paragraph{Multimodal web document datasets} The recent most performant vision and language models are trained on large sets of multimodal web documents. For instance, Flamingo \citep{Flamingo}, an 80 billion multimodal model, was trained on a mix of 2.1 billion image-text pairs, 27 million video-text pairs, and 43 million multimodal web documents. The latter called \texttt{M3W}, includes 185 million images. Similarly,  KOSMOS-1 \citep{KOSMOS} was trained on a mixture containing 71 million multimodal web documents. However, in both cases, the dataset is not publicly available, and little information is accessible as to the dataset's content, the strategies employed to create that dataset (including filtering strategies), and the quality of the resulting web documents, which ultimately hinders further research.

Concurrently to our work, the Multimodal C4 (\texttt{mmc4}) dataset \citep{MultimodalC4} was recently made accessible. It consists of 103 million multimodal web documents that include 585 million images. Although there are similarities between our datasets, it is important to highlight particular distinctions.
First, our dataset is based on more recent documents from February 2020 to February 2023, whereas \texttt{mmc4} uses documents from April 2019. 
Additionally, our filtering heuristics appear to be more comprehensive: we leverage the HTML DOM trees to filter out undesirable texts and images, whereas \texttt{mmc4} uses the HTML to find images in order to merge them with the original C4 dataset by solving a bipartite assignment problem based on a CLIP model similarities.
Last, we implement additional deduplication steps at the image, document, and paragraph levels.

\section{Creation of the Multimodal Web Document Dataset}

\begin{figure}[ht]
\centering
\resizebox{0.85\textwidth}{!}{
\input{plots/tikset_v2}
}
\caption{Overview of the steps involved in creating \texttt{OBELICS}.}
\label{fig:main_fig_creation_obelics}
\end{figure}

This section provides an overview of the critical choices of the creation and filtering process. Figure \ref{fig:main_fig_creation_obelics} gives a high-level summary of the main steps involved.
Many details are omitted from this section, and we invite the reader to refer to the appendix \ref{appendix:creation_of_the_web_doc_dataset} for completeness.

\subsection{Collecting a Large Number of HTML Files}

First, we collect a vast amount of raw web documents by considering the 25 most recent Common Crawl dumps at the time of the creation, spanning from February 2020 to January/February 2023\footnote{\url{https://commoncrawl.org/}}. We extract the main text from the documents while discarding documents with text of insufficient quality. This process results in 41.2 billion documents. 

To filter out non-English content, we apply the FastText classifier \citep{FastText} to the extracted text, which removes 63.6\% of the documents. We perform a MinHash \citep{MinHash} deduplication to remove duplicate content. Additionally, we filter out documents with significant proportions of repeated paragraphs and n-grams, following the methodology used in MassiveText \citep{Gopher}. Previous studies \citep{DeduplicationGoogle, SemDeDup} have demonstrated the prevalence of duplication in crawled data and the benefits of training on deduplicated data.

Similar to \citet{GPT3}, we employ a logistic regression classifier with hashed token frequencies to ensure high-quality text. This classifier, trained using curated datasets like Wikipedia or OpenWebText \citep{OpenWebText} as positive examples and documents sampled from Common Crawl as negative ones, is fast and effective at detecting human-written text. After these steps, we are left with 1.1 billion documents and their HTML sources from the associated Common Crawl WARC files.

\subsection{Simplifying HTML Files}

The original HTML content of a document contains a wealth of valuable information that proves highly beneficial in the process of filtering out undesirable text and images. Therefore, we prioritize pre-processing the raw HTML into simplified HTML, making the subsequent extraction of textual and visual elements more efficient.

To this aim, we devise multiple pre-processing strategies for an HTML DOM tree. By manually inspecting instances of all HTML nodes, we differentiate nodes likely to contain relevant texts or images from those that should be discarded, and we formulate specific rules for each type of node. After these pre-processing steps, the resulting simplified HTML files are more than ten times smaller and have been stripped of a large proportion of generic text (spam, ads, boilerplate template, etc.) and generic images, such as logos, while retaining the relevant content.

\subsection{Extracting Multimodal Web Documents}

In this step, we transform the simplified HTML files previously obtained into a structured web multimodal web document format. This format consists of interleaved texts and images.

We meticulously preserve the original structure of the web pages from the simplified HTML files by extracting the texts and image links while maintaining their rendering defined by the DOM tree. Given that each HTML tag denotes a distinct separation between the preceding and subsequent nodes, we leverage that information to retain line breaks and line feeds on the original page, preserving the formatting and visual rendering of the content.

We obtain 3.6 billion image links and successfully download 55\% of them (approximately 2 billion images).

\subsection{Filtering Multimodal Web Documents}

The filtering process comprises two distinct steps operating at different granularity levels. In the first step, filtering occurs at the node level for images and the paragraph level for text. This step guarantees that only high-quality and relevant images and paragraphs are retained. Each paragraph or image is evaluated based on specific criteria and may undergo modifications or be eliminated if necessary. The second step, conducted at the document level, involves deciding whether to retain or discard the output documents obtained from the first step. Most text filters used in both steps are primarily derived from \citet{ROOTS}.

\paragraph{Node-level image filtering} We discard images that are too small, excessively large or have disproportionate dimensions. We observe that these images are often indicative of low-quality or irrelevant content. To eliminate some logos and generic images, we remove images whose URLs contain one of the banned sub-strings, like \textit{logo}.

\paragraph{Paragraph-level text filtering} We apply multiple filters to text paragraphs to remove undesirable content. Specifically, paragraphs that contain an insufficient number of words are discarded. Additionally, we filter out paragraphs with high repetition ratios, excessive ratios of special characters, low ratios of stop words, low punctuation ratios, high proportions of flagged words associated with adult or inappropriate content, or excessively high perplexity scores (as measured by an n-gram language model trained on Wikipedia \citep{KenLM}). To identify boilerplate sentences or invitations to share articles on social networks, we create a list of frequently used words associated with these paragraphs and remove paragraphs containing an excessive proportion of words from this list. To further identify machine-generated content, we extract words from web-crawled documents to form a list of common words and discard documents with a low ratio of common words.

\paragraph{Document-level filtering} At the document level, we remove all documents with no or excessively high number of images. For text filters, the same filters used at the paragraph level are applied, with sometimes stricter cutoff values.

After these filtering steps, we are left with 365 million web documents and 1.4 billion images. At this step, images can be duplicated across documents.

\subsection{Responsible Filtering and Deduplication}

We take measures to minimize the amount of inappropriate content in the dataset. In particular, based on manual inspections and tool availability, we implement filters to respect data consent and remove images with pornographic content. Additionally, we also heavily deduplicate content.

\paragraph{Exclusion of opted-out images}
To respect the preferences of content creators, we remove all images for which creators explicitly opted out of AI model training. We used the Spawning API\footnote{\url{https://api.spawning.ai/spawning-api}} to verify that the images in the dataset respect the original copyright owners' choices.

\paragraph{Image deduplication based on URL}
Some images could be present across different documents. We observe that it is particularly true for browser-specific icons or common advertisements encountered during the crawling process. To address this issue, we remove all images that appear more than ten times across the entire dataset. We intentionally do not perform strict deduplication, as we notice that when an image is duplicated only a few times across different documents, the surrounding text and contextual information tend to be different. We also deduplicate images within the same document.

\paragraph{NSFW image filtering}
To reduce explicit adult content, we use an open-source NSFW classifier to remove entire documents containing pornographically classified images. We also filter out images with URLs containing banned sub-strings.

\paragraph{Document deduplication based on URL and set of images}

We complete the initial deduplication step by forming clusters of documents with the same URLs and retaining the most recent document within each cluster. We repeat this operation by forming clusters of documents containing identical sets of images.

\paragraph{Paragraph deduplication across documents of the same domain names}
To remove generic spam phrases commonly found at the end of documents, we perform paragraph-level exact deduplication within documents sharing the same domain name, resulting in the elimination of approximately 15\% of the text.

Following these filtering and deduplication steps, the final dataset contains 141 million documents and 353 million images, of which 298 million are unique. We observe that using stricter values for the filtering steps yields fewer multimodal documents, although not of higher quality. As such, we invite users who are interested in manipulating a smaller subset of \texttt{OBELICS} to start with a random subset.

\section{Analysis of \texttt{OBELICS}}\label{section:analysis_obelics}

Figure \ref{fig:viz_web_doc} provides an example showcasing an original webpage alongside the resulting multimodal web document. Extracting and filtering the multimodal document is non-trivial as it requires carefully removing undesirable information on the left, top, and bottom of the page, such as menus and navigation bars. We provide other examples at \url{https://huggingface.co/spaces/HuggingFaceM4/obelics\_visualization} and in Figures \ref{fig:web_doc_ex_1}, \ref{fig:web_doc_ex_2} and \ref{fig:web_doc_ex_3}.

Given the scale of \texttt{OBELICS}, it would be prohibitive to describe its content exhaustively. Instead, we provide high-level statistics and analyses that shed light on the dataset's properties.

\subsection{General Statistics}

\begin{figure}[htbp]
  \begin{minipage}{0.61\textwidth}
    \small
    \centering
    \captionsetup{type=table}
    \captionsetup{font=small}
    \begin{tabular}{ L{1.35cm} R{1cm} R{1.00cm} R{0.92cm} R{1cm} C{0.8cm} }
     \toprule
     Dataset&Images&\% unique images&Docs&Tokens&Open\rule[-10pt]{0pt}{0pt}\\
     \hline
     \rule{0pt}{12pt}\texttt{KOSMOS-1}&-&-&71M&-&\xmark\\
     \texttt{M3W}&185M&-&43M&-&\xmark\\
     \texttt{mmc4-ff}&385M&60.6\%&79M&34B&\cmark\\
     \texttt{mmc4}&\textbf{585M}&-&103M&43B&\cmark\\
     \texttt{OBELICS}&353M&\textbf{84.3}\%&\textbf{141M}&\textbf{115B}&\cmark\\
     \bottomrule
    \end{tabular}
    \vspace{0.5em}
    \caption{General statistics of \texttt{OBELICS} and the current largest alternatives.}
    \label{tab:general_stats}
  \end{minipage}
  \hfill
  \begin{minipage}{0.38\textwidth}
  \centering
  \captionsetup{font=small}
  \begin{tikzpicture}
    \begin{axis}[
        xlabel style={align=center}, xlabel={\footnotesize max $\#$ of images in doc},
        ylabel style={align=center}, ylabel={\footnotesize $y = \%$ of images\\\footnotesize belonging to a doc\\\footnotesize with at most $x$ images},
        xmin=0,
        xmax=100,
        ymin=0,
        ymax=100,
        grid=both,
        width=0.83\textwidth,
        height=0.83\textwidth,
        legend style={at={(1,0)}, anchor=south east}
      ]
      
      \addplot[blue, mark=*, mark size=1] coordinates {
        (0, 0.0)
        (1, 8.37)
        (2, 14.82)
        (3, 20.54)
        (4, 26.19)
        (5, 31.14)
        (6, 35.69)
        (7, 39.88)
        (8, 43.72)
        (9, 47.13)
        (10, 50.41)
        (11, 53.08)
        (12, 55.74)
        (13, 57.8)
        (14, 59.91)
        (15, 61.92)
        (16, 63.63)
        (17, 65.19)
        (18, 66.82)
        (19, 68.18)
        (20, 69.49)
        (21, 70.66)
        (22, 71.85)
        (23, 72.85)
        (24, 73.96)
        (25, 75.0)
        (26, 75.86)
        (27, 76.73)
        (28, 77.51)
        (29, 78.43)
        (30, 79.23)
        (31, 79.93)
        (32, 80.55)
        (33, 81.18)
        (34, 81.87)
        (35, 82.41)
        (36, 83.0)
        (37, 83.56)
        (38, 83.96)
        (39, 84.31)
        (40, 84.8)
        (41, 85.19)
        (42, 85.65)
        (43, 85.99)
        (44, 86.41)
        (45, 86.79)
        (46, 87.2)
        (47, 87.61)
        (48, 87.84)
        (49, 88.1)
        (50, 88.41)
        (51, 88.66)
        (52, 88.86)
        (53, 89.23)
        (54, 89.51)
        (55, 89.76)
        (56, 89.93)
        (57, 90.09)
        (58, 90.27)
        (59, 90.54)
        (60, 90.79)
        (61, 91.01)
        (62, 91.2)
        (63, 91.38)
        (64, 91.57)
        (65, 91.71)
        (66, 91.9)
        (67, 92.06)
        (68, 92.18)
        (69, 92.29)
        (70, 92.41)
        (71, 92.53)
        (72, 92.63)
        (73, 92.74)
        (74, 92.89)
        (75, 92.98)
        (76, 93.03)
        (77, 93.17)
        (78, 93.25)
        (79, 93.4)
        (80, 93.46)
        (81, 93.53)
        (82, 93.65)
        (83, 93.8)
        (84, 93.88)
        (85, 94.04)
        (86, 94.18)
        (87, 94.25)
        (88, 94.33)
        (89, 94.36)
        (90, 94.47)
        (91, 94.55)
        (92, 94.64)
        (93, 94.68)
        (94, 94.78)
        (95, 94.82)
        (96, 94.88)
        (97, 94.92)
        (98, 94.94)
        (99, 94.96)
        (100, 95.0)
      };
      \addlegendentry{\texttt{mmc4}}

      \addplot[red, mark=*, mark size=1] coordinates {
      (0, 0.0)
        (1, 21.86)
        (2, 33.03)
        (3, 42.86)
        (4, 51.73)
        (5, 58.73)
        (6, 64.38)
        (7, 68.82)
        (8, 72.59)
        (9, 76.18)
        (10, 79.24)
        (11, 81.83)
        (12, 84.01)
        (13, 85.69)
        (14, 87.38)
        (15, 88.85)
        (16, 90.22)
        (17, 91.5)
        (18, 92.63)
        (19, 93.59)
        (20, 94.62)
        (21, 95.4)
        (22, 96.17)
        (23, 96.78)
        (24, 97.36)
        (25, 97.92)
        (26, 98.41)
        (27, 98.9)
        (28, 99.23)
        (29, 99.67)
        (30, 100.0)
        (31, 100.0)
        (32, 100.0)
        (33, 100.0)
        (34, 100.0)
        (35, 100.0)
        (36, 100.0)
        (37, 100.0)
        (38, 100.0)
        (39, 100.0)
        (40, 100.0)
        (41, 100.0)
        (42, 100.0)
        (43, 100.0)
        (44, 100.0)
        (45, 100.0)
        (46, 100.0)
        (47, 100.0)
        (48, 100.0)
        (49, 100.0)
        (50, 100.0)
        (51, 100.0)
        (52, 100.0)
        (53, 100.0)
        (54, 100.0)
        (55, 100.0)
        (56, 100.0)
        (57, 100.0)
        (58, 100.0)
        (59, 100.0)
        (60, 100.0)
        (61, 100.0)
        (62, 100.0)
        (63, 100.0)
        (64, 100.0)
        (65, 100.0)
        (66, 100.0)
        (67, 100.0)
        (68, 100.0)
        (69, 100.0)
        (70, 100.0)
        (71, 100.0)
        (72, 100.0)
        (73, 100.0)
        (74, 100.0)
        (75, 100.0)
        (76, 100.0)
        (77, 100.0)
        (78, 100.0)
        (79, 100.0)
        (80, 100.0)
        (81, 100.0)
        (82, 100.0)
        (83, 100.0)
        (84, 100.0)
        (85, 100.0)
        (86, 100.0)
        (87, 100.0)
        (88, 100.0)
        (89, 100.0)
        (90, 100.0)
        (91, 100.0)
        (92, 100.0)
        (93, 100.0)
        (94, 100.0)
        (95, 100.0)
        (96, 100.0)
        (97, 100.0)
        (98, 100.0)
        (99, 100.0)
        (100, 100.0)
      };
      \addlegendentry{\texttt{OBELICS}}
    \end{axis}
  \end{tikzpicture}
  \caption{Distribution of images.}
  \label{fig:distribution_images}
  \end{minipage}
\end{figure}

Table \ref{tab:general_stats} compares \texttt{OBELICS} against the largest existing alternatives. \texttt{mmc4-ff} is the \texttt{mmc4} dataset with fewer faces. Our dataset has the highest number of unique documents and total tokens while containing a huge number of images.

It is worth mentioning that we have fewer images than \texttt{mmc4} \citep{MultimodalC4}. This discrepancy can be attributed to two reasons. First, our analysis reveals that \texttt{mmc4} contains many duplicated images, with only 60.6\% being unique compared to 84.3\% for \texttt{OBELICS}. We found that images duplicated multiple times often indicate spam or unrelated generic content. Second, \texttt{mmc4} does not limit the number of images within a document. As a result, the distribution of images across documents is highly uneven, with a substantial portion of them concentrated in documents with excessive image counts (see Figure \ref{fig:distribution_images}). The images in these documents are often unrelated to each other and exhibit spam or advertisement content. Moreover, these documents often have little text, making them unsuitable for learning the alignment between text and images (see an example in Figure \ref{fig:bad_doc_many_images}).

Figure \ref{fig:heatmap_num_tokens_num_images} shows the joint distribution of a number of tokens and a number of images in \texttt{OBELICS}. Although we limit the number of images in a document to 30, we cut the plot at 6 images for clarity. The documents of \texttt{OBELICS} contain a median number of images of 1 and a median number of tokens of 677.

\input{plots/plot_perplexity}

\paragraph{Perplexity analysis} To assess the quality of our text in comparison to reference datasets used for training large language models, we leverage an n-gram language model trained on Wikipedia \citep{KenLM, ROOTS}. This allows us to compute perplexity scores for 100,000 documents from each dataset. Lower perplexity scores indicate a higher resemblance to Wikipedia documents. Figure \ref{fig:distrib_perplexity_scores} displays the distributions of these scores. Our results demonstrate that the texts in \texttt{OBELICS} have a significantly lower average perplexity compared to the texts in c4 \citep{T5_C4}, \texttt{mmc4} \citep{MultimodalC4}, and OSCAR \citep{OSCAR}. Furthermore, our distribution aligns closely with the one from The Pile \citep{Pile}, which was thoughtfully curated from diverse, high-quality sources.

\subsection{Topic Modeling}

Similar to \cite{MultimodalC4}, we employ a Latent Dirichlet Allocation (LDA) \citep{LDA} to understand the diversity of the dataset.
The LDA gives us insights into the distribution of topics in the dataset, along with estimated proportions and frequently associated words.
Table \ref{tab:topic_modeling_20} and \ref{tab:topic_modeling_200}  present the results of the LDA with respectively 20 and 200 topics,
offering both a high-level and a more granular analysis of the dataset's content. We observe that the dataset covers topics ranging from Politics to Health by way of Music. Additionally, we compute the most frequent domains and show that news sites are systematically the most represented (Table \ref{tab:top_domains}).

\subsection{Qualitative Assessment of Dataset Samples}

We manually inspect 250 documents from \texttt{OBELICS} to verify the dataset's quality and asses the risks contained in the dataset. We focus on the images' content in relation to the text since it's the core addition compared to a language modeling dataset.


80\% of documents have photo images, while 29\% have graphic images (drawings, cartoons, etc.).
90\% of the documents have all images clearly related to the text content.
30\% of documents have images containing at least one written word, and 5\% of documents have images that are structured text (slides, tables, scanned documents, etc.), which can help models learn OCR capabilities.
7\% of documents have content (images or text) that hasn't been captured by cleaning filters (non-English text, spam or advertisement, etc.).
46\% of documents contain images with faces (portraits or group photos).
No obvious Personally Identifiable Information (PII) texts were found, except for public personalities and people mentioned in news articles.
No NSFW images were found.
Only 3\% of documents contain images with watermarks, and 2\% have images with logos.

\section{Validating the Viability of \texttt{OBELICS}}

To confirm the viability of our dataset, we first show that vision and language models trained on our multimodal web documents outperform the same models trained on image-text pairs on various multimodal benchmarks. Following that, we demonstrate the effectiveness of \texttt{OBELICS} as an alternative to closed datasets by training models of different sizes on par with closed-source models.

\paragraph{Model details} We follow the Flamingo \citep{Flamingo} architecture closely: we combine two frozen unimodal backbones - LLaMA \citep{LLaMA} for the language model, and OpenClip \footnote{\url{https://laion.ai/blog/large-openclip/}} for the vision encoder - add learnable cross-attention Transformer blocks to connect the language and vision blocks. For multimodal web documents, we feed the model sequences corresponding to the succession of text paragraphs and images. For image-text pairs, we form the training sequences by packing images with their captions. The images are encoded with the vision encoder and vision hidden states are pooled with Transformer Perceiver blocks and then fused into the text sequence through the cross-attention blocks. The training objective is the standard next token prediction. For more details, we refer to the original paper.
\newline
Following \citet{Flamingo}, we evaluate our models on a series of multimodal benchmarks spanning visual question answering (VQAv2 \citep{VQA}, OKVQA \citep{okvqa}, TextVQA \citep{textvqa}, VizWiz \citep{VizWiz}), visual dialogs (VisDial \citep{VisDial}), hateful speech detection (HatefulMeme \citep{hatefulmeme}), image captioning (COCO \citep{coco}, Flickr30k \citep{flickr30k}), and OCR (IIIT5k \citep{iiit5k}).
\newline
Additional details about the architecture, the training, the compute and the evaluation are present in Appendix \ref{sec:training_details}.

\input{plots/plot_data_laws} 

\paragraph{Training on different mixture of data} Figure \ref{fig:data_scale_law} shows the result of the first experiment, which consists in training 9B-parameter models on different mixture of data. Training on multimodal web documents allows reaching the same performance using an order of magnitude fewer images than training on image-text pairs, even though the images from the two datasets come from Common Crawl. This underlines the benefit of having longer text contexts for training multimodal models. Moreover, the model trained on multimodal web documents performs better on average. This is particularly striking on visual question-answering benchmarks on which the model trained on image-text pairs slowly degrades through the training. We note, however, that the model trained on image-text pairs has a slight advantage performance-wise in captioning, classification, and OCR tasks (see more details in Appendix \ref{sec:add_exp_results}). We hypothesize that this is due to the nature of image-text pairs: captions can be seen as fuzzy class labels. Last, similarly to \cite{Flamingo}, we observe that combining the two types of datasets leads to increased performance for a given number of images, tokens, or training compute.

\begin{table}[]
\centering
\begin{tabular}{cccccccccc}
\toprule
& Shot & \rotatebox[origin=c]{90}{COCO} & \rotatebox[origin=c]{90}{Flickr30k} & \rotatebox[origin=c]{90}{VQAv2} & \rotatebox[origin=c]{90}{OKVQA} & \rotatebox[origin=c]{90}{TextVQA} & \rotatebox[origin=c]{90}{VizWiz} & \rotatebox[origin=c]{90}{VisDial} & \rotatebox[origin=c]{90}{HatefulMemes} \\ \midrule
Flamingo-9B & \multirow{3}{*}{0} & 79.4 & 61.5 & 51.8 & 44.7 & 31.8 & 22.8 & 48.0 & 57.0 \\
OpenFlamingo-9B &  & 79.5 & 59.5 & 52.7 & 37.8 & 24.2 & 27.5 & - & 51.6 \\
IDEFICS-9B &  & 46.0 & 27.3 & 50.9 & 38.4 & 25.9 & 35.5 & 48.7 & 51.8 \\ \midrule
Flamingo-9B & \multirow{3}{*}{4} & 93.1 & 72.6 & 56.3 & 49.3 & \textbf{33.6} & 34.9 & 50.4 & 62.7 \\
OpenFlamingo-9B &  & 89.0 & 65.8 & 54.8 & 40.1 & 28.2 & 34.1 & - & 54.0 \\
IDEFICS-9B &  & 93.0 & 59.7 & 55.4 & 45.4 & 27.6 & 36.9 & 47.9 & 50.7 \\ \midrule
Flamingo-9B & \multirow{3}{*}{8} & 99.0 & \textbf{73.4} & 58.0 & 50.0 & \textbf{33.6} & 39.4 & 51.2 & 63.9 \\
OpenFlamingo-9B &  & 96.3 & 62.9 & 54.8 & 41.1 & 29.1 & 38.5 & - & 54.7 \\
IDEFICS-9B &  & 97.0 & 61.9 & 56.4 & 47.7 & 27.5 & 40.4 & 47.6 & 51.1 \\ \midrule
Flamingo-9B & \multirow{3}{*}{16} & 102.2 & 72.7 & 59.4 & 50.8 & 33.5 & 43.0 & \textbf{51.3} & \textbf{64.5} \\
OpenFlamingo-9B &  & 98.8 & 62.8 & 54.3 & 42.7 & 27.3 & 42.5 & - & 53.9 \\
IDEFICS-9B &  & 99.7 & 64.5 & 57.0 & 48.4 & 27.9 & 42.6 & - & 50.1 \\ \midrule
Flamingo-9B & \multirow{3}{*}{32} & \textbf{106.3} & 72.8 & \textbf{60.4} & \textbf{51.0} & 32.6 & \textbf{44.0} & 50.4 & 63.5 \\
OpenFlamingo-9B &  & 99.5 & 61.3 & 53.3 & 42.4 & 23.8 & \textbf{44.0} & - & 53.8 \\
IDEFICS-9B &  & 98.0 & 64.3 & 57.9 & 49.6 & 28.3 & 43.7 & - & 49.8 \\
\midrule
\midrule
Flamingo & \multirow{2}{*}{0} & 84.3 & 67.2 & 56.3 & 50.6 & 35.0 & 31.6 & 52.0 & 46.4 \\
IDEFICS &  & 91.8 & 53.7 & 60.0 & 45.2 & 30.9 & 36.0 & 48.9 & 60.6 \\ \midrule
Flamingo & \multirow{2}{*}{4} & 103.2 & 75.1 & 63.1 & 57.4 & 36.5 & 39.6 & 55.6 & 68.6 \\
IDEFICS &  & 110.3 & 73.7 & 63.6 & 52.4 & 34.4 & 40.4 & 48.4 & 57.8 \\ \midrule
Flamingo & \multirow{2}{*}{8} & 108.8 & 78.2 & 65.6 & 57.5 & 37.3 & 44.8 & 56.4 & \textbf{70.0} \\
IDEFICS &  & 114.3 & 76.6 & 64.8 & 55.1 & 35.7 & 46.1 & 47.9 & 58.2 \\ \midrule
Flamingo & \multirow{2}{*}{16} & 110.5 & 78.9 & 66.8 & \textbf{57.8} & 37.6 & 48.4 & \textbf{56.8} & \textbf{70.0} \\
IDEFICS &  & \textbf{116.6} & 80.1 & 65.4 & 56.8 & 36.3 & 48.3 & - & 57.8 \\ \midrule
Flamingo & \multirow{2}{*}{32} & 113.8 & 75.4 & \textbf{67.6} & \textbf{57.8} & \textbf{37.9} & 49.8 & 55.6 & \textbf{70.0} \\
IDEFICS &  & \textbf{116.6} & \textbf{81.1} & 65.9 & \textbf{57.8} & 36.7 & \textbf{50.0} & - & 52.5 \\
\bottomrule
\end{tabular}
\vspace{0.5em}
\caption{Performance of \texttt{IDEFICS} against OpenFlamingo and Flamingo. The evaluations were done with random in-context examples, and in an open-ended setting for VQA tasks.\\
(Task, Metric, Query split): (COCO, CIDEr, test), (Flickr30k, CIDEr, test (Karpathy)), (VQAv2, VQA acc., testdev), (OKVQA, VQA acc., val), (TextVQA, VQA acc., val), (VizWiz, VQA acc., testdev), (VisDial, NDCG, val), (HatefulMemes, ROC-AUC, test seen).}
\label{table:perf_flamingo}
\end{table}

\paragraph{Models trained on \texttt{OBELICS} achieve competitive performance at different scales} Following these insights, we show that \texttt{OBELICS} is a viable open alternative to other datasets. We train \texttt{IDEFICS}, an 80 billion parameters Flamingo-like model on a mixture of image-text pairs from LAION \citep{LAION-5B}, openly accessible captioning datasets \citep{FLAVA}, \texttt{OBELICS} and multimodal web documents obtained from Wikipedia using a similar extraction strategy. We also train a smaller version of 9 billion parameters, \texttt{IDEFICS-9B}. We compare these models against OpenFlamingo v2 \citep{OpenFlamingo} and Flamingo of the same sizes and trained on a similar mixture of multimodal web documents and image-text pairs. We report the results in Table \ref{table:perf_flamingo}.

\texttt{IDEFICS} is often on par with Flamingo on various multimodal benchmarks. Out of the 8 evaluation tasks, with 32 in-context examples, it either performs better or obtain the same result as Flamingo on 4 of them. At the 9 billion parameter scale, we are still behind Flamingo-9B. However, it is important to highlight that we outperform OpenFlamingo-9B, which was trained on \texttt{mmc4}, in terms of aggregated performance. We achieved a score of 56.5, compared to their score of 55.8, by selecting the best performance across all numbers of in-context examples for each task. This highlights the advantages of \texttt{OBELICS} as an open alternative to a multimodal web document dataset.

\section{Conclusion}

With the goal of supporting open-source large multimodal models, we introduce \texttt{OBELICS}, an open web-scale collection of filtered interleaved multimodal web documents based on Common Crawl snapshots. We document a collection and filtering process that balances the scale and removal of undesirable texts and images while addressing some of the well-documented ethical concerns of large-scale multimodal datasets, notably data consent and pornographic content. To demonstrate the usefulness of models trained on multimodal documents, we train \texttt{IDEFICS} on \texttt{OBELICS} and show that it is a viable alternative to closed datasets. Open datasets of multimodal documents with scale, quality, and diversity of sources can help support the ability to train competitive open models.

\newpage

\begin{ack}
The authors were granted access to the HPC resources of the Institut du développement et des ressources en informatique scientifique (IDRIS) du Centre national de la recherche scientifique (CNRS) under the allocation 2022-A0121013450 made by Grand équipement national de calcul intensif (GENCI). The initial development of the dataset was done on Jean-Zay cluster of IDRIS, and we thank the IDRIS team for their responsive support throughout the project, in particular Rémi Lacroix. We thank Guillaume Salou for setting up the virtual machines used to download the images of our dataset, and Sebastian Nagel for his valuable assistance in providing insights on Common Crawl. We thank Yacine Jernite and Daniel van Strien for conducting a bias analysis of the models trained on \texttt{OBELICS}.
\end{ack}


\nocite{*}
\bibliographystyle{chicago}
\bibliography{sections/references_bib}

\newpage

\section*{Checklist}

\begin{enumerate}

\item For all authors...
\begin{enumerate}
  \item Do the main claims made in the abstract and introduction accurately reflect the paper's contributions and scope?
    \answerYes{}
  \item Did you describe the limitations of your work?
    \answerYes{See Section \ref{section:analysis_obelics}.}
  \item Did you discuss any potential negative societal impacts of your work?
    \answerYes{We think that the release of such a dataset strikes a constructive trade-off between the risks associated with datasets built on top of crawled web pages (for instance, the presence of images with faces, the potential of PII in texts, offensive, insulting or threatening, etc.) with the future works that a dataset of such scale, quality and thoughtful filtering can enable. We further discuss these points in \ref{sec:ethics}.
    }
  \item Have you read the ethics review guidelines and ensured that your paper conforms to them?
    \answerYes{We read the ethics review guidelines and tried our best to match the expectations. Our content is extracted from publicly available websites at the time of the web crawl. Given the size of our dataset, it would be prohibitive to get the explicit consent of the authors of these websites. Instead, we respect the choice of content creators by removing opted-out images.
    Such a strategy cannot be exhaustive and we remain available for content creators to opt-out of the dataset.}
\end{enumerate}

\item If you are including theoretical results...
\begin{enumerate}
  \item Did you state the full set of assumptions of all theoretical results?
    \answerNA{}
	\item Did you include complete proofs of all theoretical results?
    \answerNA{}
\end{enumerate}

\item If you ran experiments (e.g. for benchmarks)...
\begin{enumerate}
  \item Did you include the code, data, and instructions needed to reproduce the main experimental results (either in the supplemental material or as a URL)?
    \answerYes{We will release the code used for the creation of the model and its training, along with the model itself.}
  \item Did you specify all the training details (e.g., data splits, hyperparameters, how they were chosen)?
    \answerYes{See Appendix \ref{sec:training_details}.}
	\item Did you report error bars (e.g., with respect to the random seed after running experiments multiple times)?
    \answerNA{}
	\item Did you include the total amount of compute and the type of resources used (e.g., type of GPUs, internal cluster, or cloud provider)?
    \answerYes{See Appendix \ref{sec:training_details}.}
\end{enumerate}

\item If you are using existing assets (e.g., code, data, models) or curating/releasing new assets...
\begin{enumerate}
  \item If your work uses existing assets, did you cite the creators?
    \answerYes{We mentioned the libraries we used.}
  \item Did you mention the license of the assets?
    \answerYes{We only used open-source libraries.}
  \item Did you include any new assets either in the supplemental material or as a URL?
    \answerNA{}
  \item Did you discuss whether and how consent was obtained from people whose data you're using/curating?
    \answerYes{See the ethics review guidelines part.}
  \item Did you discuss whether the data you are using/curating contains personally identifiable information or offensive content?
    \answerYes{The dataset we are releasing is built from publicly accessible websites. As such, there is no content in our dataset that hasn't been publicly visible on the web at some point.  Similarly, the dataset might contain texts or images that can be considered offensive, insulting, or threatening, as such data is
    prevalent on the web. We took measures to remove pornographic content and low-quality texts as much as possible. We did not take additional intentional measures to remove personal information. A manual inspection of 250 random samples reveals that there isn't obvious personally identifiable information (excluding celebrities and people mentioned in news articles), although it is likely that the dataset contains some.}
\end{enumerate}

\item If you used crowdsourcing or conducted research with human subjects...
\begin{enumerate}
  \item Did you include the full text of instructions given to participants and screenshots, if applicable?
    \answerNA{}
  \item Did you describe any potential participant risks, with links to Institutional Review Board (IRB) approvals, if applicable?
    \answerNA{}
  \item Did you include the estimated hourly wage paid to participants and the total amount spent on participant compensation?
    \answerNA{}
\end{enumerate}

\end{enumerate}

\newpage

\appendix

\section{Appendix}

\subsection{Creation of the Multimodal Web Document Dataset}\label{appendix:creation_of_the_web_doc_dataset}

\subsubsection{Collecting of a Large Number of HTML Files}

Our data collection process begins by considering the 25 most recent Common Crawl\footnote{https://commoncrawl.org/} dumps available at the time of dataset creation. It contains webpages spanning from February 2020 to January/February 2023.
We use a modified version of \texttt{readability-lxml}\footnote{https://github.com/buriy/python-readability} to extract the main text from the pages, discarding any pages that contain text of excessively high perplexity. This process yields a total of 41.2 billion documents.

\paragraph{Selection of English content} To identify non-English content, we apply the FastText classifier \citep{FastText} to the extracted text, effectively filtering out 63.6\% of the documents.

\paragraph{Early text deduplication} Often, a set of URLs is crawled repeatedly across different Common Crawl snapshots. However, the content of these websites may vary as web administrators make changes over time. Hence, at this stage, we refrain from deduplicating documents based on their URLs. Instead, we perform MinHash \citep{MinHash} deduplication with 16 hashes calculated over 5-grams. To further refine the data, we eliminate documents containing substantial proportions of repeated paragraphs and n-grams, employing the methodology described in MassiveText \citep{Gopher}. \citep{DeduplicationGoogle, SemDeDup} show that crawled data often contains a significant amount of duplication, and training on deduplicated data can improve performance. 

\paragraph{Quality classification} We employ a logistic regression classifier with hashed token frequencies to only retain pages containing human-written text, similar to \citet{GPT3}. The classifier is trained using documents from curated datasets, such as Wikipedia and OpenWebText \citep{OpenWebText}, as positive examples, and documents sampled from Common Crawl as negative examples. For simplicity, we use a threshold of 0.5 for the probability that a document comes from a curated corpus, which acts as an indicator that a document is human-written.

Following these steps, we obtain 1.1 billion documents and their HTML sources from the associated Common Crawl WARC files.

\subsubsection{Simplifying HTML Files}

The original HTML content of a document contains a wealth of valuable information that proves highly beneficial in the process of filtering out undesirable text and images. Therefore, we prioritize pre-processing the raw HTML into simplified HTML, making the subsequent extraction of textual and visual elements more efficient. For this purpose, we use the library \texttt{selectolax}\footnote{https://github.com/rushter/selectolax} that facilitates efficient parsing of HTML files and creates corresponding DOM trees.

\paragraph{DOM Tree cleaning strategies} To simplify the DOM trees, we employ several cleaning strategies. Firstly, we convert tags that indicate line breaks (such as \texttt{<br>}) into actual line breaks. Multiple consecutive line breaks and spaces are condensed into a single instance. Additionally, HTML comments are removed from the DOM trees. Furthermore, we implement recursive processes to eliminate empty leaves and unnest nodes. When a parent node lacks attached text and has only one child, the child node replaces the parent node in the DOM hierarchy. We repeat these operations after removing some nodes, and describe this process in the following paragraphs.

\paragraph{Tag unwrapping} This operation involves removing unnecessary styling applied to displayed text by unwrapping a predefined set of tags given below. By applying this procedure, tags such as \texttt{<i>example</i>} are transformed into \texttt{example}, eliminating the associated styling elements.

The following tags are unwrapped during the processing of HTML files: \texttt{a}, \texttt{abbr}, \texttt{acronym}, \texttt{b}, \texttt{bdi}, \texttt{bdo}, \texttt{big}, \texttt{cite}, \texttt{code}, \texttt{data}, \texttt{dfn}, \texttt{em}, \texttt{font}, \texttt{i}, \texttt{ins}, \texttt{kbd}, \texttt{mark}, \texttt{q}, \texttt{s}, \texttt{samp}, \texttt{shadow}, \texttt{small}, \texttt{span}, \texttt{strike}, \texttt{strong}, \texttt{sub}, \texttt{sup}, \texttt{time}, \texttt{tt}, \texttt{u}, \texttt{var}, \texttt{wbr}.

\paragraph{Node removal} Following the previous step, we conduct a manual inspection of practical examples encompassing all existing HTML tags. Based on our findings, we establish a curated list that outlines the tags we intend to retain. Any nodes within the HTML DOM tree with tags not included in this list are subsequently removed. We specifically retain tags that define the document structure (e.g., \texttt{p} or \texttt{h}) and tags associated with media elements (e.g., \texttt{img}). However, we opt to remove tags that typically consist of logos, generic content, or spam (e.g., \texttt{header}), as well as tags that often contain noisy text related to website navigation (e.g., \texttt{li}), or text that poses challenges in terms of linearization (e.g., \texttt{table}).

We retain the following tags during the processing of HTML files, as they define the document's structure: \texttt{address}, \texttt{article}, \texttt{aside}, \texttt{blink}, \texttt{blockquote}, \texttt{body}, \texttt{br}, \texttt{caption}, \texttt{center}, \texttt{dd}, \texttt{dl}, \texttt{dt}, \texttt{div}, \texttt{figcaption}, \texttt{h}, \texttt{h1}, \texttt{h2}, \texttt{h3}, \texttt{h4}, \texttt{h5}, \texttt{h6}, \texttt{hgroup}, \texttt{html}, \texttt{legend}, \texttt{main}, \texttt{marquee}, \texttt{ol}, \texttt{p}, \texttt{section}, \texttt{summary}, \texttt{title}, \texttt{ul}.
Additionally, we also preserve the following tags that define media elements: \texttt{audio}, \texttt{embed}, \texttt{figure}, \texttt{iframe}, \texttt{img}, \texttt{object}, \texttt{picture}, \texttt{video}.
Furthermore, we keep the \texttt{source} tag as it may contain an interesting attribute.

\paragraph{Modification of specific nodes} We then specifically target some \texttt{<div>} nodes that contain \texttt{footer}, \texttt{header}, \texttt{navigation}, \texttt{nav}, \texttt{navbar}, or \texttt{menu} as ID or \texttt{date} as attribute, as well as CSS rules that possess \texttt{footer} or \texttt{site-info} as class. These nodes typically contain website navigation content or article dates and are therefore removed. Additionally, we observe that the presence of a CSS rule with the class \texttt{more-link} often indicates a distinct shift in topic within the webpage, resembling the start of a new document. To account for this, we replace these nodes with the text \texttt{END\_OF\_DOCUMENT\_TOKEN\_TO\_BE\_REPLACED}, which we replace by an end-of-sentence (EOS) token during training.

With these processing steps, we reduce the size of the HTML files by more than 10 on average while preserving the interesting content.

\subsubsection{Extracting Multimodal Web Documents}

In this section, we begin with the simplified HTML files obtained from the previous section. Our objective is to transform these files into a structured web document format, which is a sequence of interleaved texts and images.

\paragraph{Preservation of the original structure of the web pages}

During the extraction process, we meticulously preserve the original structure of the web pages from the simplified HTML files. We extract the texts and image links while maintaining their order of appearance in the DOM tree. Each HTML tag denotes a distinct separation between the preceding and subsequent nodes and we retain any line breaks and line feeds that are present in the original page, preserving the formatting and visual rendering of the content.

\paragraph{Image downloading}

To download the images, we use the \texttt{img2dataset} \citep{img2dataset} library. We attempt to download a massive collection of 3.6 billion images, of which 55\% (approximately 2 billion images) were successfully downloaded. For that, we employ 20 virtual machines. This distributed approach allow us to complete the operation within a few days.

\subsubsection{Filtering Multimodal Web Documents}

The filtering process consists of two steps, targeting different levels of granularity. In the first step, filtering occurs at the node level for images and at the paragraph level (separated by line breaks) for text. We evaluate each paragraph or image and we potentially modify or remove these based on specific criteria. The second step, conducted at the document level, involves deciding whether to retain or discard the output documents from the first step. The majority of the filters for text we use for both steps were adapted from \citet{ROOTS}.

\paragraph{Node-level image filtering} We discard images with formats other than \texttt{jpg}, \texttt{png} or \texttt{webp}, with a side length below 150 pixels or exceeding 20,000 pixels, as well as those with an aspect ratio greater than 2 or less than 1/2. These criteria help exclude images that are too small, excessively large, or have disproportionate dimensions, which are often indicative of low-quality or irrelevant content. To eliminate some logos and generic images, as in \citep{MultimodalC4}, we remove images whose URL contains one of the sub-strings \textit{logo}, \textit{button}, \textit{icon}, \textit{plugin} or \textit{widget}.

\paragraph{Paragraph-level text filtering}
Regarding text paragraphs, we apply a series of filters to remove undesirable or irrelevant content. We discard paragraphs with fewer than 4 words, as they typically contain insufficient information to be considered meaningful. Additionally, we remove paragraphs with a high repetition ratio, indicating potential spam content, and those with an excessive ratio of special characters, often associated with irrelevant or low-quality text. 
\newline
Furthermore, we filter out paragraphs with a low ratio of stop words, as it is often indicative of machine-generated or nonsensical content. Similarly, we exclude paragraphs with a low punctuation ratio, as they typically indicate poor-quality texts. We also consider the flagged word ratio, removing paragraphs with a high proportion of flagged words associated with adult or inappropriate content. We also use KenLM \citep{KenLM} models trained on Wikipedia to filter out paragraphs with excessively high perplexity scores.
\newline
To minimize spam, one approach is to identify generic sentences or invitations to share articles on social networks commonly found at the end of documents. We create a list of frequently used words associated with these paragraphs and then filter out paragraphs that contain an excessive proportion of words from this list.
\newline
To augment our ability to identify non-human-generated content, we consider a subset of 10 million documents from OSCAR \citep{OSCAR}, a web-crawled corpus. We extract the words from these documents, removed punctuations, converted them to lowercase, and retain only the words occurring at least twice, which we refer to as common words. We filter out paragraphs with a too low common word ratio.
\newline
The detail of the cutoff values for all text filters at the paragraph level is present in Table \ref{tab:cutoffs_text_filters}.

By applying these node-level and paragraph-level filters, we ensure that only high-quality and relevant images and paragraphs are retained for further processing and analysis.

\begin{table}[h]
\centering
\begin{tabular}{ p{5.5cm} p{1.5cm} p{2.5cm} p{2.5cm} }
 \hline
 \rule{0pt}{12pt}Metric &Cutoff type &Cutoff value (paragraph-level)&Cutoff value (document-level)\rule[-10pt]{0pt}{0pt}\\
 \hline
 \rule{0pt}{12pt}Number of words &min &4 &10\\
 Number of words &max &1,000 &2,000\\
 Character repetition ratio &max &0.1 &0.1\\
 Word repetition ratio &max &0.1 &0.2\\
 Special character ratio &max &0.3 &0.275\\
 Stop word ratio &min &0.3 &0.35\\
 Flagged word ratio &max &0.01 &0.01\\
 Punctuation ratio &min &0.001 &0.03\\
 Spam word ratio &max &0.12 &0.12\\
 Common word ratio &min &0.8 &0.9\\
 Language identification prediction score &min &0.8 &0.8\\
 Perplexity score &max &1500 &1500\rule[-10pt]{0pt}{0pt}\\
 \hline
\end{tabular}
\vspace{0.5em}
\caption{Cutoff values for text filters at paragraph and document levels. A 'min' (or 'max') cutoff indicates that any paragraph or document, depending on the level, with a value for the considered metric strictly below (or above) the cutoff value is removed.}
\label{tab:cutoffs_text_filters}
\end{table}

\paragraph{Document-level filtering}
For document-level filtering, we start by removing all documents with no images or with more than 30 images. We have found that when there are too many images in a document, they are often not related to each other, and are more likely to be considered as spam.
\newline
For text filters, we use the same filters as for filtering at paragraph level. Since we are at the document level, the filter metrics are more precise, and we can typically set stricter cutoff values while limiting the number of false positives. The cutoff values used are also present in Table \ref{tab:cutoffs_text_filters}. 

After these filtering steps, we obtained 365 million web documents and 1.4 billion images (potentially duplicated in different documents at this stage).

\subsubsection{Additional Filtering and Deduplication Steps}

\paragraph{Exclusion of opted-out images}
To respect the preferences of content creators, we remove all images for which creators explicitly opted out of AI model training. We used the Spawning API\footnote{\url{https://api.spawning.ai/spawning-api}} to verify that the images in the dataset respect the original copyright owners' choices. This step had a small impact on the overall dataset, by removing only 0.047\% of the images.

\paragraph{Image deduplication based on URL}
Prior to this step, it is possible for the same image to be present in multiple documents under the same URL. However, we observe that the distribution of image occurrences was highly skewed, with the majority of images appearing only once, while a small subset of images appeared hundreds of thousands of times. Upon closer examination, we notice that these frequently occurring images are predominantly comprised of common advertisements encountered during the crawling process, browser-specific icons, and similar elements. To address this issue, we remove all images that appear more than 10 times across the entire dataset. This approach significantly reduces the presence of unwanted images. We intentionally do not perform strict deduplication, as we observe that when an image is duplicated only a few times across different documents, the surrounding text and contextual information tend to vary. These diverse contexts associated with the duplicated image could be beneficial for the training of a model. We also deduplicate images within the same document.

\paragraph{NSFW image removal}
We use an open-source NSFW classifier\footnote{\url{https://github.com/GantMan/nsfw_model}} to reduce the proportion of explicit adult content within our dataset. We carefully choose a cutoff that reduces as much as possible the proportion of false positives. Indeed, if favoring precision to recall may seem to be a good idea to remove as much undesirable content as possible, it hurts diversity. An analysis of false positives shows that in many cases, simple portrait photos of women are classified as pornographic, which is not the case for men. People of color are also more often misclassified. We remove the entire document when a pornographically classified image is found in the document. In addition, we also remove all images whose URLs contain the sub-strings \textit{porn}, \textit{sex} or \textit{xxx}. We remove approximately 1\% of the documents with this filter. Note that many pornographic documents have been previously removed by the filter on flagged words.

\paragraph{Document deduplication based on URL}
Since we consider many Common Crawl dumps, it is possible that several documents may be associated with the same URL, despite the initial deduplication efforts. Recognizing the inherent similarity among these documents, we opt to retain only the most recent document for each common URL.

\paragraph{Document deduplication based on set of images}
It is possible that documents with different URLs and domain names are very similar and have not been removed by the first deduplication, for instance, news articles copied and pasted multiple times across various sources. To mitigate this, we form groups of documents with an identical set of images, and we keep only the most recent document for each group.  

\paragraph{Paragraph deduplication across documents of the same domain names}
To eliminate generic spam phrases commonly found at the end of documents, such as "Share on Facebook," "Post a comment," or "Accept the cookies," we implement a paragraph-level deduplication process within documents sharing the same domain name. This approach aims to enhance the quality of the text by removing redundant and repetitive content. For each domain name, we identify paragraphs that appear at least three times in an identical manner across associated documents. These repetitive paragraphs are subsequently removed from the documents, resulting in the elimination of approximately 15\% of the text present in the web documents.

After all these steps, the final dataset contains 141 million documents and 353 million images, of which 298 million are unique.

We observe that using stricter values for the filtering steps yields fewer multimodal documents, although not of higher quality. As such, we invite users who are interested in manipulating a smaller subset of \texttt{OBELICS} to start with a random subset.


\newpage

\subsection{Analysis of \texttt{OBELICS}}

\subsubsection{Examples of Multimodal Web Documents}

\begin{figure}[H]
\includegraphics[width=0.66\textwidth]{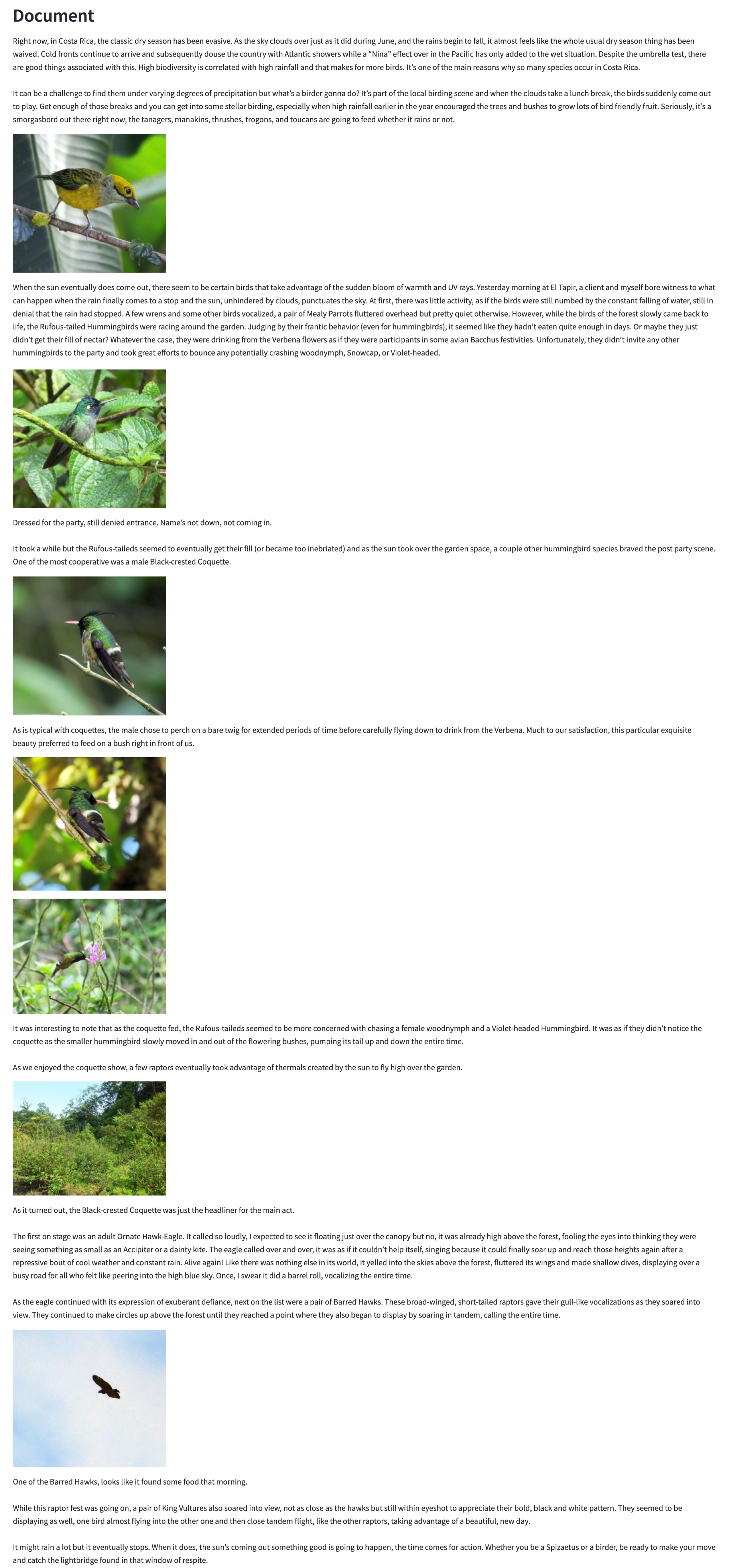}
  \caption{Example of a document in \texttt{OBELICS}.\\ \scriptsize From \url{http://birdingcraft.com/wordpress/2018/01/23/what-happens-with-birding-in-costa-rica-when-the-rain-stops/}}
  \label{fig:web_doc_ex_1}
\end{figure}

\newpage

\begin{figure}[H]
\includegraphics[width=0.71\textwidth]{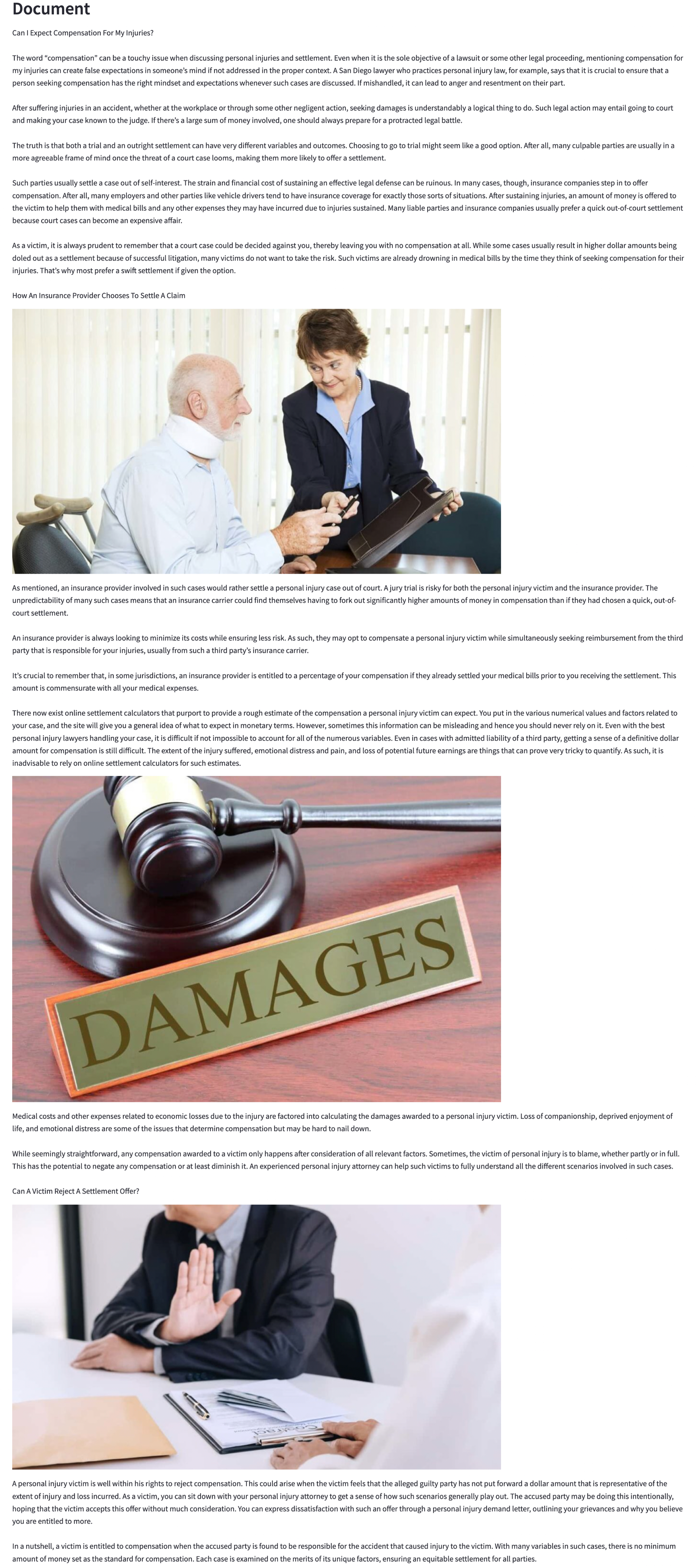}
  \caption{Example of a document in \texttt{OBELICS}.\\ \scriptsize From \url{https://www.halt.org/can-i-expect-compensation-for-my-injuries/}}
  \label{fig:web_doc_ex_2}
\end{figure}

\newpage

\begin{figure}[H]
\includegraphics[width=0.85\textwidth]{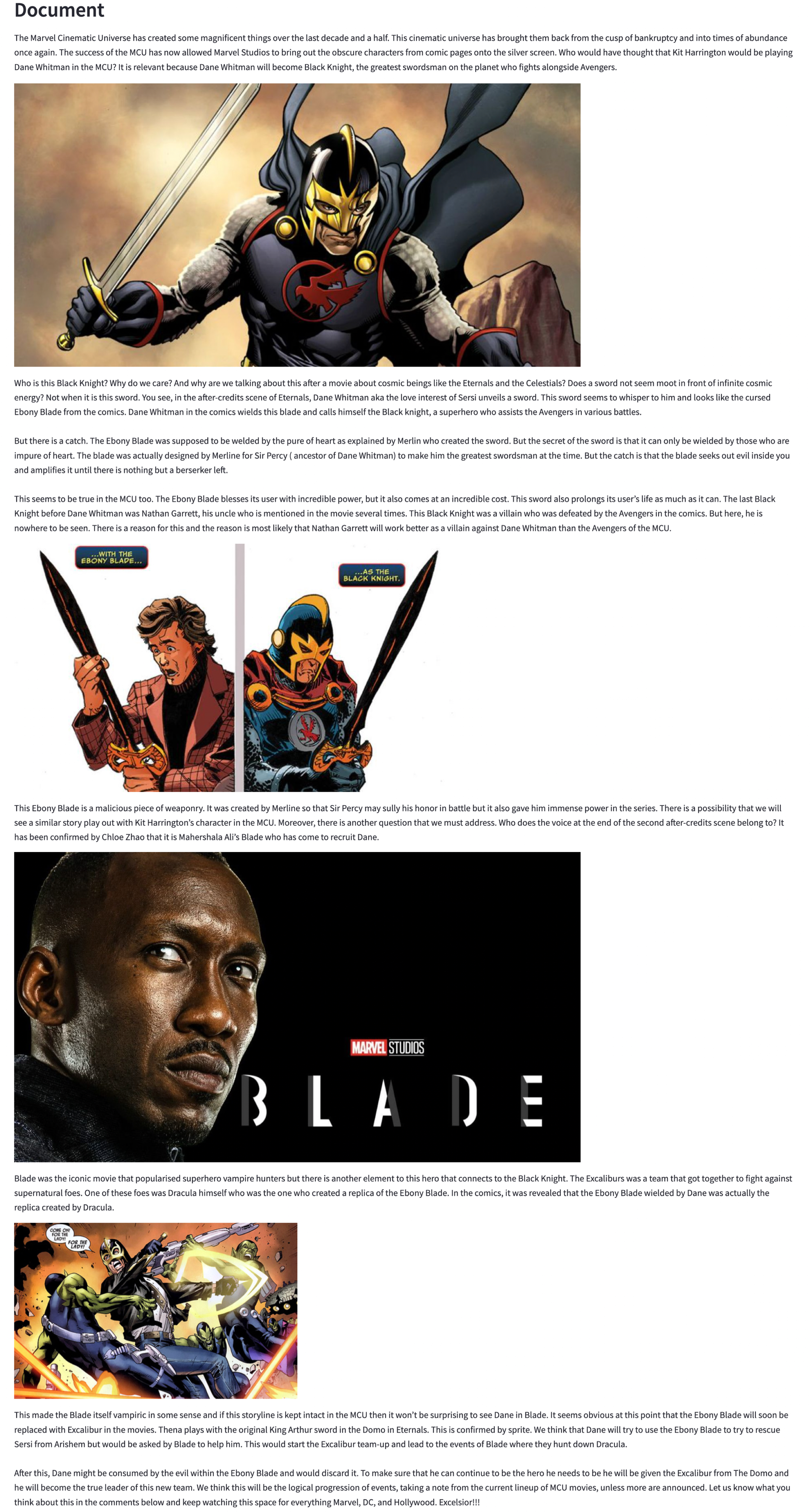}
  \caption{Example of a document in \texttt{OBELICS}.\\ \scriptsize From \url{https://www.quirkybyte.com/blog/2021/11/how-dane-whitman-will-become-black-knight-kit-harringtons-character-explained/}}
  \label{fig:web_doc_ex_3}
\end{figure}

\newpage

\subsubsection{Unwanted Document Containing Many Images}

\begin{figure}[H]
\includegraphics[width=0.14\textwidth]{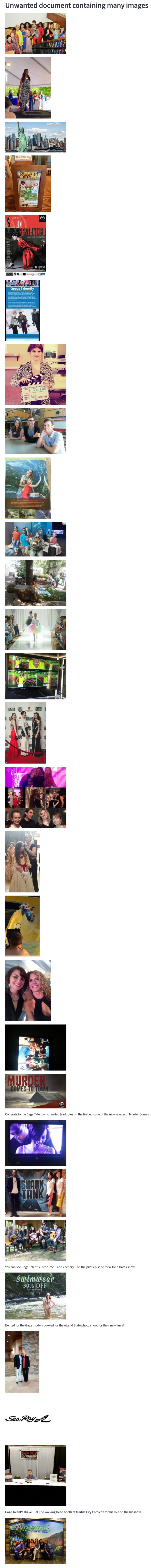}
  \caption{Undesirable document containing many images. Text is only present in small proportions, and the relation between the images is not always clear.}
  \label{fig:bad_doc_many_images}
\end{figure}

\newpage

\subsubsection{Top 100 Domains}

\begin{longtable}{p{1cm} p{5.5cm} p{1.75cm}}
 \hline
 \rule{0pt}{12pt}Rank&Domain name&Number of documents\rule[-10pt]{0pt}{0pt}\\
 \hline
 \rule{0pt}{12pt}1&www.dailymail.co.uk&434,498\\
2&en.wikipedia.org&155,258\\
3&nypost.com&141,494\\
4&www.thestar.com&138,224\\
5&sputniknews.com&133,695\\
6&www.rediff.com&133,233\\
7&www.theepochtimes.com&132,539\\
8&www.fool.com&125,220\\
9&www.businessinsider.com.au&123,841\\
10&www.bustle.com&122,581\\
11&www.dailysabah.com&120,029\\
12&www.firstpost.com&119,642\\
13&www.irishtimes.com&118,329\\
14&theathletic.com&101,982\\
15&www.news.com.au&98,339\\
16&www.indiatimes.com&98,197\\
17&www.theglobeandmail.com&92,805\\
18&tvtropes.org&92,104\\
19&www.dailydot.com&91,034\\
20&mashable.com&88,310\\
21&observer.com&87,336\\
22&www.cbsnews.com&86,759\\
23&www.rappler.com&86,554\\
24&www.tmz.com&84,472\\
25&www.salon.com&84,420\\
26&www.modernghana.com&83,918\\
27&www.foxnews.com&83,002\\
28&www.huffpost.com&81,701\\
29&www.ndtv.com&81,549\\
30&www.thisismoney.co.uk&80,930\\
31&www.famousbirthdays.com&78,931\\
32&www.engadget.com&76,817\\
33&www.rnz.co.nz&76,327\\
34&www.metro.us&75,627\\
35&www.patheos.com&75,003\\
36&www.news24.com&73,883\\
37&www.thestar.com.my&73,265\\
38&www.dw.com&72,774\\
39&www.npr.org&71,939\\
40&koreajoongangdaily.joins.com&71,091\\
41&peoplesdaily.pdnews.cn&71,048\\
42&pagesix.com&70,602\\
43&www.thenigerianvoice.com&70,470\\
44&wikimili.com&69,928\\
45&www.indiebound.org&67,986\\
46&www.cricketcountry.com&66,605\\
47&expressdigest.com&64,250\\
48&www.capitalfm.co.ke&64,163\\
49&www.bizpacreview.com&64,157\\
50&www.wionews.com&63,797\\
51&profootballtalk.nbcsports.com&63,532\\
52&jamaica-gleaner.com&63,137\\
53&www.rte.ie&63,074\\
54&www.aspentimes.com&62,552\\
55&kids.kiddle.co&62,419\\
56&english.alarabiya.net&60,368\\
57&www.jellypages.com&59,381\\
58&people.com&59,293\\
59&muse.jhu.edu&59,061\\
60&www.geeky-gadgets.com&58,975\\
61&www.khaleejtimes.com&58,851\\
62&www.nbcsports.com&57,922\\
63&en.topwar.ru&56,723\\
64&www.thewrap.com&56,146\\
65&www.outlookindia.com&55,752\\
66&www.celebdirtylaundry.com&55,618\\
67&time.com&55,527\\
68&www.dailystar.co.uk&55,503\\
69&www.legit.ng&55,395\\
70&www.thehansindia.com&55,109\\
71&www.bbc.co.uk&55,015\\
72&newsinfo.inquirer.net&54,927\\
73&nesn.com&54,756\\
74&www.tellerreport.com&53,939\\
75&www.rawstory.com&53,676\\
76&www.thestatesman.com&53,286\\
77&wccftech.com&52,510\\
78&forward.com&51,969\\
79&nationalinterest.org&51,851\\
80&www.pearltrees.com&50,933\\
81&www.contactmusic.com&50,284\\
82&www.tweaktown.com&50,138\\
83&www.destructoid.com&50,081\\
84&www.publishersweekly.com&49,735\\
85&www.cbs58.com&49,680\\
86&www.markedbyteachers.com&48,994\\
87&www.caughtoffside.com&48,857\\
88&www.islamicinvitationturkey.com&48,721\\
89&dailyhive.com&48,447\\
90&www.aljazeera.com&47,393\\
91&www.bbc.com&47,349\\
92&worldbulletin.dunyabulteni.net&47,300\\
93&www.romper.com&47,115\\
94&www.catchnews.com&47,025\\
95&www.odt.co.nz&46,712\\
96&www.jewishpress.com&46,688\\
97&www.irishcentral.com&46,629\\
98&techcrunch.com&46,539\\
99&www.nhl.com&46,247\\
100&www.tuko.co.ke&46,106\rule[-10pt]{0pt}{0pt}\\
\hline
\caption{\rule{0pt}{12pt}Ranking of the 100 domains with the highest number of associated documents in \texttt{OBELICS}.}
\label{tab:top_domains}
\end{longtable}

\subsubsection{Topic Modeling with 20 Topics}

\begin{longtable}{p{.15\textwidth} p{.10\textwidth} p{.65\textwidth}}
\hline
\rule{0pt}{12pt}Concept&Ratio&Related words\rule[-10pt]{0pt}{0pt}\\ \hline
\rule{0pt}{12pt}Justice&5.16\%&said, police, people, year, according, court, case, told, news, man, two, death, also, one, old, investigation, found, fire, officers\\
Politics&6.35\%&said, state, government, would, president, trump, law, court, party, public, new, election, states, political, federal, house, people, also, bill\\
Family&5.24\%&family, one, day, back, life, time, home, would, old, said, years, like, two, love, mother, children, first, man, went\\
Music&5.23\%&music, album, band, song, new, songs, show, also, first, sound, rock, one, musical, year, released, live, festival, record, track\\
Climate&3.46\%&water, energy, climate, species, also, earth, space, one, used, gas, use, solar, natural, power, carbon, years, change, system, may\\
Business&7.12\%&year, company, million, market, said, new, business, companies, per, also, billion, percent, price, financial, money, industry, years, growth, according\\
Sports&3.75\%&game, season, team, first, year, two, said, three, play, last, games, one, win, second, points, coach, back, players, four\\
Sports (2nd)&5.67\%&team, first, year, season, league, last, two, club, world, race, one, game, win, time, back, players, match, second, final\\
Automotive&4.18\%&new, car, also, design, one, power, cars, two, model, use, used, system, camera, first, speed, engine, high, vehicle, battery\\
Cinema&7.36\%&film, story, series, movie, book, new, show, one, also, characters, character, first, world, star, films, love, best, life, man\\
War&4.26\%&war, country, said, military, countries, russia, world, russian, government, united, international, people, states, president, also, security, israel, army, forces\\
Gaming&5.77\%&game, use, also, new, games, data, one, users, app, online, using, video, google, players, play, time, used, information, content\\
Health&3.0\%&health, also, may, medical, patients, disease, study, people, treatment, cancer, body, use, drug, research, risk, brain, care, virus, cases\\
Food&2.08\%&food, also, one, beer, like, eat, made, wine, restaurant, make, coffee, meat, well, used, tea, sugar, use, water, taste\\
Urban&4.62\%&city, area, new, park, one, building, town, road, also, north, day, around, river, island, south, place, along, local, two\\
Existence&5.23\%&one, people, god, life, world, women, many, even, human, may, like, way, men, often, would, man, also, social, power, must\\
Asia&1.61\%&india, indian, also, china, said, chinese, government, minister, pakistan, country, delhi, kong, hong, people, singh, two, khan, sri, asia\\
History&4.24\%&book, art, first, history, years, new, century, work, one, books, also, church, american, world, time, museum, english, known\\
Education&5.11\%&school, said, students, work, university, new, community, also, people, years, year, education, program, women, working, support, college, children, project\\
Other&10.56\%&like, one, get, would, time, people, really, know, even, think, much, good, going, way, see, could, make, want, things, something\rule[-10pt]{0pt}{0pt}\\
\hline
\caption{\rule{0pt}{12pt}LDA with 20 topics, trained on 100,000 random web documents. A concept for each topic is derived from the related words.}
\label{tab:topic_modeling_20}
\end{longtable}

\subsubsection{Topic Modeling with 200 Topics}

\begin{longtable}{p{.25\textwidth} p{.10\textwidth} p{.55\textwidth}}
\hline
\rule{0pt}{12pt}Concept&Ratio&Related words\rule[-10pt]{0pt}{0pt}\\ \hline
\rule{0pt}{12pt}Celebrity Relationships&0.52\%&star, fans, show, love, instagram, couple, together, shared, relationship, revealed, year, kim, charlie, told, actress, pete, new, former, old, lisa\\
Music Industry&1.47\%&band, music, song, album, songs, rock, tour, live, singer, show, record, country, bands, released, stage, one, love, played, pop\\
Racial Diversity&0.26\%&black, white, people, race, african, american, racial, community, racism, gay, racist, americans, diversity, lgbtq, justice, color, lgbt, gender, discrimination, queer\\
Language Usage&0.17\%&language, english, word, words, name, languages, use, used, text, names, letter, letters, meaning, translation, writing, spoken, speech, speaking, speak, term\\
Team Spirit&0.38\%&said, get, team, good, really, going, lot, year, think, got, great, like, last, back, well, play, time, guys, big, hard\\
News Media&0.28\%&news, media, radio, fox, press, magazine, journalists, television, journalism, story, newspaper, editor, journalist, coverage, times, broadcast, interview, daily, podcast, show\\
European Culture&0.04\%&van, dutch, netherlands, tattoo, amsterdam, belgium, portugal, belgian, der, tattoos, portuguese, bulgaria, sofia, holland, bulgarian, lisbon, santos, europe, tulip, brussels\\
European Nations&0.19\%&european, germany, german, europe, berlin, sweden, poland, greece, also, countries, swedish, polish, czech, denmark, norway, austria, greek, hungary, finland\\
Film Industry&1.29\%&film, movie, films, director, movies, best, actor, hollywood, documentary, cinema, role, screen, story, directed, production, actors, also, oscar, award\\
Australian Achievements&0.12\%&australia, australian, new, zealand, sydney, award, melbourne, awards, year, victoria, queensland, south, nsw, brisbane, australians, best, won, auckland, prize\\
Culinary Delights&0.88\%&cream, recipe, cheese, make, chocolate, made, bread, add, taste, ice, butter, sauce, cake, sugar, cook, food, salt, milk, sweet\\
Life and Death&0.4\%&death, one, people, life, world, dead, even, lives, many, die, died, lost, killed, still, never, man, end, left, day, hope\\
Spiritual Philosophy&0.2\%&philosophy, spiritual, buddhist, religion, religious, yoga, buddha, meditation, buddhism, tibetan, guru, book, practice, knowledge, thought, mind, life, modern, texts, tradition\\
Cultural Histories&0.13\%&jewish, jews, indigenous, native, holocaust, rabbi, tribe, people, indian, community, peoples, tribal, israel, tribes, anti, culture, land, camp, history, torah\\
Personal Development&0.07\%&says, people, explains, like, new, adds, get, work, want, also, tells, lot, say, year, years, really, working, part, wants, help\\
Royal Families&0.23\%&king, prince, royal, queen, princess, charles, henry, elizabeth, duke, harry, palace, meghan, family, william, anne, castle, kate, lady, diana, edward\\
Daily News&0.19\%&said, week, friday, monday, wednesday, according, tuesday, thursday, news, last, day, told, sunday, saturday, reported, statement, days, morning, hours\\
Creative Projects&0.19\%&project, design, work, working, projects, creative, create, idea, team, process, also, ideas, new, make, designer, created, started, concept, worked, wanted\\
Legal Investigations&0.6\%&investigation, information, former, report, fbi, department, office, according, documents, evidence, public, intelligence, government, claims, allegations, corruption, fraud, alleged, officials, federal\\
Medical Procedures&0.19\%&surgery, skin, pain, treatment, cancer, procedure, patients, teeth, bone, patient, surgical, injury, eye, hair, tissue, surgeon, tooth, breast, honey, medical\\
Athletic Competitions&0.46\%&olympic, sports, world, athletes, games, sport, olympics, gold, team, medal, NUMm, event, won, year, championships, competition, athlete, time, first\\
Historical Artifacts&0.62\%&ancient, century, NUMth, history, temple, stone, roman, years, one, city, also, greek, found, known, built, old, site, time, today\\
Literary Works&0.87\%&book, books, read, story, author, novel, writing, reading, series, stories, first, written, fiction, published, readers, characters, world, one, write, new\\
Time Progression&0.73\%&one, year, years, last, still, could, even, time, big, new, two, much, like, back, next, would, since, another, well, already\\
Everyday Life&0.2\%&day, time, sleep, night, home, hours, room, water, house, bed, days, morning, work, get, every, food, hour, two, camp, minutes\\
Colorful Nature&0.16\%&color, tea, dark, white, green, flowers, skin, like, black, flower, colors, blue, rose, leaves, light, pink, also, red, used, golden\\
Automotive Industry&1.21\%&car, cars, engine, vehicle, new, vehicles, model, electric, ford, drive, also, wheel, rear, speed, driving, toyota, motor, front, power\\
American Cities&0.11\%&new, york, california, city, san, los, angeles, francisco, chicago, jersey, state, times, diego, brooklyn, center, santa, bay, seattle, county\\
Political Movements&0.57\%&political, people, power, party, government, right, america, politics, anti, war, state, world, left, free, nation, democracy, american, country, media, system\\
Mythical Creatures&0.12\%&bear, wolf, dragon, snake, bears, lion, like, tiger, monster, wild, human, wolves, animals, snakes, cave, creatures, giant, humans, hunter, dragons\\
Asian Cultures&0.09\%&north, korea, harry, kim, korean, potter, south, jon, thrones, jong, pyongyang, stewart, nuclear, ron, warner, hogwarts, house, game, colbert, peninsula\\
Data Modeling&0.31\%&data, model, number, value, using, numbers, function, used, models, values, two, example, method, figure, one, set, problem, object, line\\
Romantic Stories&1.34\%&story, love, life, girl, one, new, woman, find, young, man, finds, characters, father, friend, two, character, family, romance, secret, series\\
Medical Research&0.41\%&cancer, cells, cell, dna, disease, gene, human, patients, genetic, immune, protein, treatment, genes, bacteria, researchers, diseases, research, proteins, study, clinical\\
Fitness and Training&0.21\%&running, race, run, training, marathon, fitness, miles, exercise, bike, mile, runners, NUMk, course, gym, finish, cycling, yoga, half, runner\\
Personal Perspectives&1.43\%&like, people, think, really, would, know, going, get, see, one, lot, things, something, time, want, way, much, thing, say, could\\
Gastronomy Scene&0.44\%&food, restaurant, coffee, bar, restaurants, menu, chef, chicken, pizza, meal, kitchen, dishes, dinner, eat, dining, burger, table, meals, served, like\\
Labor Rights&0.29\%&workers, work, employees, job, jobs, union, pay, labor, working, employment, insurance, employers, wage, employee, company, paid, worker, labour, staff, business\\
Competitive Sports&0.75\%&game, second, goal, first, ball, half, back, minutes, win, lead, two, points, score, minute, final, match, side, three, time\\
Public Events&0.71\%&year, event, festival, christmas, day, events, NUMth, show, night, tickets, special, holiday, party, live, celebrate, held, also, place, saturday\\
Digital Marketing&0.37\%&digital, content, marketing, media, brand, advertising, platform, online, campaign, ads, business, industry, social, new, users, platforms, brands, companies, internet, consumers\\
Public Safety&0.24\%&safety, report, action, letter, statement, said, incident, ban, made, public, actions, claims, reported, according, response, taken, complaints, following, take, serious\\
French Heritage&0.1\%&french, france, paris, jean, saint, les, des, pierre, dame, marie, europe, macron, notre, louis, european, michel, jamaica, jacques, emmanuel\\
Eastern European Politics&0.38\%&russian, russia, ukraine, ukrainian, moscow, putin, soviet, state, vladimir, war, azerbaijan, country, armenian, armenia, president, russians, union, sanctions, region\\
Horror Entertainment&0.58\%&movie, story, horror, characters, character, film, action, one, plot, ghost, scene, evil, movies, like, series, original, genre, dark, scenes, first\\
Political Campaigns&1.25\%&trump, president, election, vote, campaign, obama, party, biden, house, donald, political, republican, presidential, voters, democratic, democrats, candidate, clinton, candidates, white\\
Indian Cinema&0.64\%&film, khan, actor, also, movie, bollywood, films, kapoor, indian, actress, seen, role, singh, india, release, hindi, kumar, directed, hai, salman\\
Corporate Leadership&0.82\%&years, board, director, president, team, business, leadership, work, executive, also, chief, role, member, management, service, experience, served, staff, working\\
Law Enforcement&1.94\%&police, said, officers, man, officer, arrested, year, old, incident, two, found, according, investigation, killed, department, shot, scene, vehicle, suspect\\
Football Clubs&1.26\%&club, league, season, united, premier, players, city, football, chelsea, team, arsenal, player, manchester, liverpool, game, side, back, last, games\\
Essential Skills&0.84\%&get, make, need, one, also, time, best, want, many, use, may, take, find, like, even, help, way, good, people, much\\
Artistic Expression&0.75\%&art, museum, artist, work, artists, exhibition, painting, works, gallery, arts, paintings, collection, artistic, drawing, new, show, contemporary, painted, artwork\\
American Regions&0.22\%&state, county, texas, florida, north, south, michigan, ohio, carolina, states, virginia, west, georgia, center, university, washington, colorado, iowa, arizona\\
Industrial Production&0.28\%&production, company, industry, mining, manufacturing, gold, mine, port, supply, project, companies, factory, industrial, plant, steel, products, equipment, coal, goods\\
Global Affairs&0.36\%&world, countries, international, united, trade, china, states, global, country, foreign, europe, region, asia, economic, european, nations, south, india, east\\
Government Affairs&1.26\%&minister, government, said, meeting, party, president, prime, would, members, committee, council, parliament, also, general, decision, agreement, political, secretary, national, commission\\
Software Development&0.67\%&code, use, file, using, software, version, files, windows, run, server, application, web, source, open, user, system, new, linux, install\\
UK Happenings&0.22\%&london, british, england, britain, centre, brexit, bbc, wales, labour, west, manchester, johnson, north, programme, south, across, may, year, east\\
Real Estate Market&0.16\%&property, housing, estate, home, real, homes, house, rent, properties, market, land, mortgage, rental, sale, houses, price, owner, buyers, sales, units\\
Fashion Trends&0.43\%&fashion, hair, wearing, dress, wear, look, style, clothing, clothes, black, wore, designer, beauty, shirt, women, also, made, show, costume, new\\
Gaming Culture&0.38\%&game, cards, card, games, play, players, poker, player, casino, online, gambling, win, deck, playing, betting, lottery, bet, slot, chess, played\\
Famous Personalities&0.04\%&bond, kelly, martin, daniel, peter, doctor, tony, johnny, parker, sean, evans, frank, andy, ian, lucas, dave, reynolds, spy, emily, amber\\
Wildlife Conservation&0.61\%&species, birds, bird, animals, fish, found, animal, also, wild, wildlife, eggs, habitat, large, food, like, small, humans, insects, many, endangered\\
Pandemic Responses&0.94\%&covid, pandemic, health, people, virus, coronavirus, vaccine, cases, said, spread, outbreak, public, lockdown, vaccines, government, new, disease, vaccination, deaths\\
Popular Names&0.11\%&john, michael, david, paul, jones, james, johnson, mike, jim, steve, robert, two, bob, davis, moore, allen, brian, mark, one\\
Christian Theology&0.45\%&god, jesus, christ, bible, christian, church, faith, lord, people, gospel, paul, christians, john, prayer, word, biblical, kingdom, pastor, moses\\
Sports&0.77\%&season, team, game, nba, games, basketball, players, player, play, coach, league, hockey, points, teams, nhl, played, first, star, year\\
Cybersecurity&0.63\%&data, security, network, internet, cloud, information, access, technology, services, service, NUMg, software, computer, systems, networks, cyber, devices, users, attacks, use\\
Business/Finance&0.78\%&company, business, companies, market, industry, investment, investors, capital, tech, firm, ceo, based, technology, billion, businesses, group, million, financial, growth\\
Professional Wrestling&0.18\%&wwe, ring, wrestling, match, rick, randy, champion, title, wrestler, vince, show, fans, wrestlers, owens, tag, baker, triple, shane, raw, cody\\
Japanese Culture/Tech&0.15\%&anime, musk, japanese, tesla, manga, series, elon, japan, ninja, episode, samurai, kai, characters, demon, karate, character, also, dragon, arc, tokyo\\
Scottish Personalities&0.03\%&brown, scotland, scottish, gordon, glasgow, celtic, perry, walker, murray, graham, letter, edinburgh, cover, campbell, watson, thomas, also, well, neil, henderson\\
Streaming Media&0.12\%&video, youtube, videos, live, watch, channel, streaming, audio, content, stream, channels, footage, shows, online, also, NUMk, recording, watching, clip, one\\
Christianity&0.36\%&church, catholic, pope, religious, christian, churches, bishop, francis, faith, holy, priest, saint, mass, vatican, religion, pastor, christ, parish, christians\\
Smartphone Technology&0.83\%&phone, apple, samsung, iphone, pro, smartphone, device, galaxy, camera, also, display, battery, new, sNUM, screen, NUMgb, phones, NUMg, android\\
Urban Development&0.78\%&city, project, area, council, residents, community, park, town, street, public, local, cities, new, development, mayor, urban, construction, district, building\\
Sociocultural Issues&0.39\%&social, culture, society, cultural, people, political, different, moral, identity, important, values, issues, often, public, role, many, way, community, understanding, view\\
Common Male Names&0.03\%&smith, jack, tom, ben, adam, alex, kevin, richard, simon, holmes, billy, bell, oliver, harvey, jake, collins, burke, baldwin, joel, aaron\\
Combat Sports&0.49\%&fight, title, tennis, champion, ufc, round, world, boxing, fighter, one, win, open, martial, first, match, mma, fighters, fighting, career\\
Indian Politics&0.64\%&india, indian, state, delhi, government, also, minister, bjp, said, modi, singh, chief, congress, crore, pradesh, mumbai, gandhi, lakh, hindu\\
Military History&0.25\%&war, world, battle, empire, british, army, history, german, peace, great, military, wars, end, conflict, power, two, land, forces, soldiers, fight\\
Internet Cartography&0.04\%&www, map, sri, http, https, maps, lanka, com, atlas, derby, tamil, lankan, html, maria, angelo, tara, colombo, org, mapping, easter\\
European Football&0.46\%&league, champions, team, goals, world, season, football, club, cup, madrid, barcelona, player, real, players, match, messi, ronaldo, liverpool, final\\
Mobile Applications&0.73\%&app, google, apple, android, users, mobile, apps, phone, new, devices, device, ios, iphone, microsoft, use, also, features, user, screen, windows\\
Korean Entertainment&0.11\%&lee, korean, korea, kim, south, park, seoul, drama, group, bts, jin, jung, first, also, members, won, woo, hyun, young, min\\
Economics&1.01\%&market, price, prices, markets, growth, inflation, economy, stock, economic, rate, rates, investors, higher, year, demand, stocks, trading, dollar, gold\\
Video Games&0.49\%&games, game, xbox, gaming, nintendo, video, play, console, playstation, mario, psNUM, one, sony, players, steam, gamers, switch, playing, titles\\
Time Indicators&0.3\%&first, years, since, time, two, NUMth, three, total, day, year, may, second, september, june, january, november, four, NUM/NUM, april\\
Science Fiction/Fantasy&0.14\%&star, wars, trek, lego, luke, figures, force, series, jedi, kirk, toy, universe, figure, new, ship, galaxy, crew, fans, space, disney\\
Music Production&1.09\%&album, sound, music, band, track, song, guitar, metal, sounds, tracks, songs, record, bass, vocals, new, release, rock, like, released, drums\\
Transportation&0.42\%&document, token, road, end, replaced, bike, traffic, driving, drivers, bus, train, driver, bridge, car, station, ride, roads, route, transport, rail\\
Personal Life&1.14\%&life, people, love, world, many, time, one, always, years, great, every, like, way, friends, never, day, work, first, hope, best\\
American History&0.6\%&american, history, NUMs, new, first, years, century, america, early, states, united, NUMth, became, world, many, one, today, time, war\\
Global Policy&0.96\%&change, climate, development, economic, government, global, policy, need, sector, world, public, new, support, economy, national, social, future, health, impact, crisis\\
South Asian Affairs&0.2\%&pakistan, afghanistan, taliban, kashmir, bangladesh, khan, india, pakistani, afghan, also, nepal, country, indian, kabul, jammu, singh, islamabad, ali, lahore, karachi\\
Sports Scores&0.83\%&game, points, first, season, two, three, win, second, four, team, lead, run, third, one, five, scored, home, games, point\\
Travel/Daily Life&1.03\%&day, time, back, get, last, one, got, good, night, next, morning, went, first, trip, week, see, around, way, little\\
Announcements&0.83\%&new, year, first, last, time, next, NUMth, month, also, release, announced, two, months, march, since, october, september, week, may\\
Online Dating&0.13\%&dating, gay, online, sites, date, site, tinder, free, men, best, matchmaking, meet, guy, hookup, guys, app, apps, relationship, singles, dates\\
Superhero Comics&0.42\%&comic, marvel, comics, man, batman, spider, superhero, character, avengers, superman, universe, hero, captain, new, heroes, fans, issue, super, characters, also\\
Space Exploration&0.31\%&space, nasa, mission, mars, drone, launch, rocket, satellite, robot, earth, robots, drones, moon, first, station, orbit, satellites, spacecraft, technology\\
Musical Performance&0.57\%&music, jazz, musical, concert, piano, orchestra, composer, musicians, classical, symphony, played, performance, playing, performed, piece, work, instruments, also, festival, instrument\\
Personal Finance&0.17\%&money, pay, card, credit, bank, cash, vegas, payment, paid, account, las, payments, fees, cost, cards, amount, buy, service, fee\\
Television Shows&0.74\%&show, series, season, episode, netflix, shows, episodes, television, comedy, watch, cast, fans, also, new, seasons, character, drama, viewers, first\\
Celebrity Culture&0.11\%&taylor, jackson, justin, swift, star, jennifer, singer, jay, tyler, cohen, nicole, spencer, also, eddie, cole, carrie, amy, lopez, bieber, casey\\
Environmental Conservation&0.32\%&water, river, land, environmental, forest, wildlife, conservation, area, natural, lake, areas, project, environment, rivers, dam, resources, forests, national, management\\
Physical/Quantum Sciences&0.35\%&water, air, chemical, used, process, material, surface, materials, quantum, temperature, high, oxygen, carbon, radiation, particles, liquid, salt, energy, pollution, chemicals\\
Astronomy&0.37\%&earth, sun, moon, planet, sky, stars, solar, star, space, light, universe, planets, telescope, years, scientists, system, galaxy, eclipse, dark\\
Islamic/Middle Eastern Culture&0.19\%&muslim, saudi, muslims, islam, islamic, arabia, egypt, arab, dubai, allah, uae, ali, middle, abu, prophet, religious, muhammad, mosque, iran, egyptian\\
Gender Issues&0.14\%&women, men, woman, female, girls, gender, male, abortion, sexual, girl, young, sex, life, equality, feminist, man, violence, ladies, rights, boys\\
Fantasy/Mythology&0.03\%&sam, lewis, max, rings, twin, troy, monkey, toy, stephen, palmer, doll, hobbit, tolkien, zeus, lord, monkeys, seth, horse, toys, witch\\
Video Game Mechanics&0.36\%&attack, damage, enemy, pokemon, use, weapon, enemies, level, also, fight, battle, attacks, players, power, weapons, ability, magic, hero, character, armor\\
MMORPG Gaming&1.16\%&game, games, players, play, new, player, world, playing, characters, gameplay, mode, character, also, story, battle, fun, experience, free, fantasy\\
Energy and Environment&0.65\%&energy, oil, gas, power, carbon, solar, fuel, emissions, electricity, climate, wind, renewable, coal, natural, green, production, industry, fossil, environmental\\
Financial Regulations&0.57\%&tax, financial, bank, government, debt, income, banks, money, taxes, budget, economy, finance, loan, pay, billion, loans, credit, economic, fund\\
US Legislation&0.75\%&state, bill, would, federal, house, senate, congress, law, legislation, act, states, governor, government, passed, public, committee, lawmakers, plan, funding\\
Subjective Experience&0.91\%&like, good, really, one, well, much, great, bit, even, little, quite, also, though, still, pretty, lot, see, get, better, would\\
Parenthood&0.16\%&children, child, kids, parents, baby, age, young, birth, parent, pregnancy, pregnant, family, families, babies, adults, mother, old, early, mothers\\
Personal Experiences&1.93\%&like, get, one, know, got, really, good, little, even, think, guy, thing, going, love, pretty, right, let, much, never, back\\
Education&0.55\%&school, students, education, schools, college, student, high, university, class, teachers, year, teacher, campus, program, learning, teaching, classes, children, grade, parents\\
Latin American Cultures&0.17\%&mexico, spanish, italian, spain, italy, san, mexican, latin, puerto, del, cuba, rico, colombia, costa, america, cuban, venezuela, juan, country\\
Technological Systems&0.68\%&system, new, technology, systems, development, also, use, time, process, high, based, performance, work, used, well, using, provide, quality, level, developed\\
Social Movements&0.6\%&rights, people, government, human, violence, protest, freedom, police, country, protests, law, civil, political, protesters, movement, state, justice, activists, right, groups\\
Surfing/Beach Culture&0.02\%&scott, ryan, wilson, joe, anderson, wave, josh, sarah, phil, surf, jackie, waves, robinson, logan, beach, ken, surfing, phoenix, duncan, gibson\\
Brazilian Culture&0.03\%&brazil, brazilian, miller, rio, phillips, paulo, portuguese, peterson, grande, são, janeiro, ivy, bolsonaro, herman, silva, state, amazon, sao, spike, hernandez\\
Literature/Poetry&0.32\%&poetry, writing, essay, writer, poem, poems, literary, literature, work, poet, book, published, writers, wrote, write, english, works, collection, written, life\\
Family Life&0.58\%&family, years, wife, home, mary, born, school, life, funeral, friends, died, church, death, service, many, member, may, mrs, passed\\
Cricket&0.47\%&cricket, india, test, match, runs, team, england, series, first, wickets, ipl, overs, game, tNUM, played, indian, ball, innings, captain\\
Canadian/Irish Affairs&0.09\%&canada, canadian, ireland, irish, toronto, ontario, vancouver, dublin, province, alberta, northern, canadians, ottawa, montreal, provincial, centre, quebec, north, trudeau\\
Music Industry&1.01\%&music, album, song, artists, artist, hip, single, hop, released, new, songs, rapper, track, video, rap, pop, release, hit, singer\\
Criminal Justice&0.6\%&prison, crime, criminal, court, charges, sexual, trial, case, jail, years, crimes, guilty, victims, murder, abuse, accused, sentence, justice, convicted\\
Academic Research&0.66\%&university, research, science, professor, institute, studies, college, scientific, school, work, study, engineering, national, international, department, students, degree, academic, center\\
Names and Dates&0.02\%&williams, hill, ross, carter, kennedy, clark, jan, nelson, jordan, stanley, rated, murphy, arthur, marshall, hudson, feb, nov, oct, mar\\
Weather Conditions&0.49\%&weather, ice, snow, mountain, winter, north, temperatures, cold, climate, south, high, lake, rain, temperature, east, west, summer, conditions, ski\\
Health and Medicine&0.54\%&blood, brain, disease, symptoms, may, heart, patients, body, treatment, also, cause, risk, pain, condition, effects, common, severe, doctor, pressure\\
Cryptocurrency&0.47\%&bitcoin, blockchain, crypto, cryptocurrency, digital, mining, ethereum, cryptocurrencies, currency, exchange, btc, market, network, tokens, users, price, nft, trading, transactions, token\\
Diet and Nutrition&0.38\%&food, diet, weight, health, body, fat, eating, foods, eat, sugar, healthy, also, high, diabetes, people, meat, protein, obesity, levels\\
Actions and Movements&0.12\%&back, get, time, take, right, move, way, next, see, start, around, keep, make, end, away, going, one, left, another, turn\\
Historic Landmarks&0.36\%&NUMth, town, village, name, william, george, century, hall, john, family, built, castle, early, house, mill, street, history, became, morris\\
Electronic Devices&0.41\%&power, light, battery, use, control, device, used, system, led, also, using, devices, high, signal, air, electrical, switch, low, sensor\\
Performing Arts&0.43\%&theatre, show, dance, stage, play, theater, performance, production, audience, musical, opera, arts, broadway, dancing, cast, performances, performing, company, ballet, shakespeare\\
Mental Health&0.26\%&mental, people, health, disorder, depression, help, self, anxiety, stress, emotional, person, life, physical, may, often, brain, also, social, autism, feel\\
Online Interaction&0.35\%&post, blog, read, comments, posted, like, would, one, see, com, please, know, article, share, site, email, comment, posts, link, page\\
Substance Usage&0.27\%&drug, drugs, cannabis, marijuana, use, cbd, medical, effects, addiction, fda, used, alcohol, cocaine, substance, prescription, heroin, treatment, products, thc, also\\
Outdoor Landscapes&0.46\%&tree, trees, trail, water, road, river, along, forest, area, around, small, park, one, near, old, wood, way, hill, across, ground\\
Colors&0.06\%&red, blue, white, green, black, yellow, color, light, flag, orange, grey, colors, gray, logo, one, pearl, hat, look, colour, two\\
Israel and Fishing&0.19\%&israel, israeli, fish, palestinian, jerusalem, fishing, gaza, palestinians, netanyahu, hamas, jewish, bank, west, palestine, state, arab, israelis, trout, salmon\\
Air Travel&0.4\%&airport, flight, aircraft, air, airlines, plane, flights, travel, airline, passengers, aviation, flying, fly, international, airports, pilot, passenger, boeing, service\\
Waste and Recycling&0.16\%&plastic, waste, made, used, use, bags, make, bag, paper, items, nike, fabric, shoes, cola, using, coca, trash, recycling, also, shoe\\
Philosophical Discourse&0.34\%&would, even, one, could, however, much, fact, yet, rather, far, though, many, well, might, perhaps, less, long, despite, may, time\\
Problems and Issues&0.16\%&could, problem, many, may, problems, due, however, issues, issue, would, even, also, cause, result, still, time, situation, damage, impact, without\\
Firearms and Malaysia&0.17\%&gun, shooting, guns, malaysia, hunting, rifle, firearms, shot, deer, weapons, shoot, weapon, malaysian, pistol, firearm, ammunition, rmNUM, hunt, buck\\
Disney and Animation&0.12\%&disney, magic, world, ray, animation, alice, walt, park, animated, fairy, ride, parks, disneyland, theme, magical, pixar, jungle, studios, orlando, characters\\
Middle Eastern Conflict&0.81\%&syria, turkey, forces, iraq, military, security, attacks, attack, killed, syrian, terrorist, turkish, war, people, state, group, isis, terrorism, terrorists, government\\
Physical Descriptions&0.48\%&eyes, like, face, could, head, hand, back, little, looked, hands, said, around, look, body, would, voice, see, away, hair, felt\\
Architecture&0.62\%&building, house, room, space, built, floor, construction, wall, buildings, new, home, design, tower, two, walls, architecture, roof, rooms, designed\\
Travel Destinations&0.94\%&city, hotel, park, one, visit, tour, world, town, place, travel, area, many, also, trip, beautiful, places, visitors, located, island\\
Computer Hardware&0.41\%&intel, performance, computer, memory, amd, core, graphics, usb, windows, laptop, drive, cpu, card, power, nvidia, hardware, gpu, processor, gaming\\
African Nations&0.17\%&africa, south, african, kenya, country, cape, uganda, rNUM, zimbabwe, continent, national, congo, africans, west, tanzania, president, town, johannesburg, rwanda, nairobi\\
Military Operations&0.37\%&military, army, war, soldiers, forces, troops, general, service, battle, soldier, commander, men, armed, corps, force, command, training, unit, guard, combat\\
Tobacco and Cookies&0.15\%&cookies, website, smoking, use, tobacco, cigarettes, buy, smoke, experience, cigar, cookie, necessary, used, ivermectin, cigarette, consent, online, may, vaping, also\\
Nigerian Politics&0.67\%&state, nigeria, said, government, nigerian, governor, president, ghana, lagos, buhari, also, nNUM, nigerians, country, national, federal, people, apc, security, abuja\\
Family Dynamics&0.54\%&family, father, mother, son, old, daughter, home, children, years, year, parents, wife, young, brother, life, dad, two, house, sister\\
Farming and Agriculture&0.4\%&plant, farmers, farm, food, plants, agriculture, garden, soil, agricultural, seeds, grow, growing, seed, crop, crops, production, farming, farms, fruit, harvest\\
Retail Industry&0.27\%&store, market, products, sales, amazon, stores, customers, price, company, business, retail, product, buy, shop, online, consumers, brand, shopping, sell, selling\\
Online Resources&0.32\%&download, information, free, page, available, online, book, edition, website, pdf, article, site, published, library, content, please, text, may, read\\
Personal Experiences&2.07\%&would, time, could, one, didn, first, back, got, went, years, came, wanted, made, started, took, never, day, wasn, thought, even\\
Theology and Morality&0.45\%&god, man, one, lord, world, life, earth, upon, power, may, spirit, human, evil, love, heaven, gods, soul, must, every, shall\\
Sports and Games&1.29\%&season, game, team, football, nfl, yards, baseball, games, players, league, coach, field, play, year, player, bowl, quarterback, teams, first\\
Asia and Pacific&0.07\%&japan, japanese, tokyo, vietnam, indonesia, pacific, hawaii, island, vietnamese, indonesian, islands, asian, also, asia, west, rice, jakarta, abe, hawaiian\\
Healthcare&0.27\%&health, care, medical, hospital, patients, doctors, healthcare, patient, treatment, services, medicine, doctor, hospitals, hiv, nursing, nurses, emergency, insurance, nurse, staff\\
Commemorations&0.21\%&day, memorial, anniversary, national, NUMth, ceremony, veterans, flag, honor, statue, cemetery, people, nation, war, country, president, service, years, monument\\
Collectibles and Auctions&0.32\%&gold, collection, silver, watch, auction, box, original, sold, coin, coins, one, made, sale, watches, design, set, edition, also, rare\\
East Asia&0.18\%&china, chinese, kong, hong, singapore, philippines, beijing, taiwan, thailand, shanghai, asia, also, thai, province, asian, country, philippine, city, manila\\
Maritime Exploration&0.4\%&sea, island, ship, boat, ocean, water, coast, beach, bay, ships, marine, islands, boats, cruise, port, waters, crew, fishing, sailing\\
Natural Disasters&0.39\%&fire, people, storm, hurricane, disaster, emergency, fires, damage, flood, earthquake, rescue, smoke, flooding, firefighters, homes, residents, burning, hit, area\\
Legal Matters&0.69\%&court, law, case, judge, legal, supreme, justice, decision, attorney, filed, trial, cases, courts, lawyer, lawyers, lawsuit, appeal, ruling, judges\\
Dimensions and Positioning&0.47\%&two, side, one, top, right, back, cut, line, use, small, used, hand, like, left, body, front, size, using, around\\
Relationships and Marriage&0.18\%&marriage, sex, relationship, married, wedding, love, couple, sexual, divorce, man, husband, wife, couples, together, woman, partner, men, one, relationships, bride\\
Community Projects&0.84\%&community, support, group, people, members, program, help, local, foundation, event, also, work, organization, part, project, together, youth, young, year\\
Photography&0.26\%&image, camera, images, photo, photos, NUMd, photography, pictures, cameras, picture, light, lens, photographer, capture, photographs, taken, shot, look, using, shoot\\
Competitive Sports&0.88\%&team, players, teams, cup, tournament, world, football, competition, final, round, golf, play, club, match, first, won, league, win, sports\\
Innovation and Science&0.57\%&world, human, new, reality, create, like, time, life, future, nature, work, experience, way, process, space, ideas, different, form, idea, science\\
Personal Opinions&1.87\%&people, know, like, think, say, even, want, make, one, something, things, someone, way, doesn, would, good, need, person, feel, never\\
Statistics&0.99\%&percent, per, year, number, according, cent, average, report, increase, years, rate, million, data, population, last, people, increased, growth, higher\\
Personal Communication&0.15\%&said, would, told, people, added, could, asked, also, going, think, want, year, last, say, saying, one, interview, make, come, according\\
Animal Companions&0.3\%&dog, dogs, cat, animals, animal, cats, horse, pet, breed, horses, pets, also, owner, bull, owners, pig, rescue, puppy, pigs, humans\\
Scientific Research&0.41\%&study, research, data, researchers, found, results, studies, risk, analysis, evidence, group, published, test, findings, based, university, likely, may, could\\
Mystery and Adventure&0.43\%&man, back, one, left, door, street, front, around, away, saw, car, went, two, night, told, heard, took, later, behind, another\\
Motor Racing&0.85\%&race, racing, team, season, track, car, races, second, first, win, championship, lap, two, driver, top, series, year, drivers, fNUM\\
International Politics&0.56\%&united, states, iran, border, trump, nuclear, president, immigration, security, country, administration, foreign, american, countries, migrants, policy, refugees, immigrants, government, washington\\
Air Defense&0.34\%&air, aircraft, force, military, navy, defense, defence, wing, fighter, missile, flying, base, naval, command, pilot, pilots, flight, forces, jet\\
Additional Information&0.62\%&within, however, additionally, stated, mentioned, one, extra, password, might, individuals, simply, time, present, actually, get, place, may, together, different\\
Financial Performance&0.62\%&million, year, billion, company, quarter, sales, revenue, per, said, share, total, according, last, first, NUMm, percent, expected, growth, reported\\
Alcohol and Beverages&0.38\%&beer, wine, drink, alcohol, brewery, drinking, wines, bottle, brewing, beers, craft, taste, brew, drinks, whisky, ale, tasting, bar, whiskey, bottles\\
Celebrity Profiles&0.66\%&also, career, born, known, years, worth, age, net, life, famous, american, became, name, first, million, started, year, appeared, actress\\
Storytelling and Narratives&1.26\%&like, life, story, world, one, time, sense, way, yet, much, work, makes, narrative, every, often, takes, moments, something, stories, piece\\
Legislation&0.78\%&law, act, rules, may, legal, laws, government, public, must, state, regulations, would, information, rule, commission, states, required, order, authority\\
Social Media&0.45\%&twitter, facebook, social, media, instagram, post, people, account, also, pic, tweet, share, news, online, posted, video, users, page, wrote, shared\\
Comparative Analysis&0.42\%&one, also, however, two, may, different, many, used, example, well, often, first, part, although, another, time, known, fact, various, number\rule[-10pt]{0pt}{0pt}\\
\hline
\caption{\rule{0pt}{12pt}LDA with 200 topics, trained on 100,000 random web documents. A concept for each topic is derived from the related words.}
\label{tab:topic_modeling_200}
\end{longtable}

\newpage

\subsection{Ethical discussion}\label{sec:ethics}
At the beginning of the project, we reflected on ethical principles\footnote{\parbox{\linewidth}{\url{https://huggingface.co/blog/ethical-charter-multimodal}}} guiding the project, including the creation of the dataset, in order to incorporate ethical values we agreed on. These values motivated the careful crafting of the content filters. For instance, we used the Spawning API to respect as much as possible the consent decisions of content creators or iterated significantly on filters around pornographic content.

Exploring large-scale corpora is often a tedious process which contributes to the lack of transparency and lack of documentation around these artifacts. With that in mind, we built an interactive visualization\footnote{\parbox{\linewidth}{\url{https://atlas.nomic.ai/map/f2fba2aa-3647-4f49-a0f3-9347daeee499/ee4a84bd-f125-4bcc-a683-1b4e231cb10f}}} of OBELICS which allows browsing through a subset (11M documents) of the dataset and navigate the different topics covered. Yet, we note that despite our efforts, OBELICS contains a small proportion of documents that are not suitable for all audiences. For instance, one might find the cluster named “Sex” which predominantly contains descriptions of pornographic movies along with pornographic images. Other clusters would contain advertising for sex workers, or reports of violent shootings. In our experience, these documents represent a small proportion of all the documents.

Due to the nature of our dataset (multimodal documents extracted from the web), OBELICS inherits the same ethical concerns of unlabeled text corpora crawled from the web: difficulty to document/inspect, presence of unintended biases, under-representation of certain demographics, etc. These concerns have been well documented for text corpora \citep{pmlr-v137-biderman20a,Bender2021OnTD}. Data audits have shed light on the some limitations and unintended biases contained in these text corpora \citep{Caswell2020LanguageII,Dodge2021DocumentingLW}. The augmentation of text corpora with interleaved images is a recent development of multimodal machine learning. We hope that our dataset along with exploration tools will serve as a solid ground for endeavors such as data audits. Existing works auditing large-scale multimodal datasets have focused on image-text pairs datasets \citep{Birhane2021MultimodalDM} and highlight how curation and filtering decisions lead to biases (including racism and misogyny) in the resulting pairs. We believe that interleaved image-text datasets will play a significant role in the development of increasingly more capable multimodal models, and having large-scale versions of these datasets that are transparent, maintained and in open-access is critical.

We also have evaluated the trained models as part of a red-teaming effort and a systematic evaluation of the generations produced by the model compared across the axis of gender and race. More specifically, the model was separately prompted to write a resume, a dating profile, and a headline about a person’s recent arrest based on their appearance. We studied the generations and analyzed the trends for each protected characteristic using FairFace \citep{FairFace} and StableBias \citep{StableBias}. The details of these evaluations and insights are made public as part of the model release. As an example, the model trained on OBELICS associates men more frequently than women with terms like “financial”, “development”, “product”, and “software”.

\subsection{Building the Model}\label{sec:training_details}

\subsubsection{Architecture Details}
We closely follow the Flamingo architecture introduced in \citet{Flamingo}. To form the model, we combine a pre-trained image encoder, a pre-trained language model, and add newly initialized parameters of the form of Perceiver blocks \citep{jaegle2021perceiver} and Transformer-based cross-attentions blocks inserted within the language model every 4 layers.

The pre-trained backbones are frozen during the training, and only the new parameters are updated along with the embeddings of additional tokens.

Following \citet{dehghani2023scaling}, we apply a layer normalization on the projected queries and keys of both the Perceiver and cross-attention blocks, which improved training stability in our early experiments. We use the RMSNorm implementation \citep{zhang-sennrich-neurips19} for the layer normalization.

\begin{table}[h]
\resizebox{\textwidth}{!}{
\begin{tabular}{cccccccc}
\toprule
 \textbf{Total} & \textbf{Trainable} & \textbf{Language Model} & \textbf{Vision Model} & \textbf{Perceiver} & \textbf{Cross-Attentions}\rule[-5pt]{0pt}{0pt} \\ \hline
\rule{0pt}{10pt}9B & 1.5B & 7B & 630M &  126M & 1.4B  \\
80B & 14B & 65B & 630M & 126M & 13.9B \\
\bottomrule
\end{tabular}
}
\vspace{0.5em}
\caption{Breakdown of model parameters. We use LLaMA \citep{LLaMA} for the language backbone and OpenCLIP (\url{https://laion.ai/blog/large-openclip/}) for the vision backbone.}
\label{tab:architecture_details}
\end{table}

\subsubsection{Training Details}

We roughly use the same set hyper-parameters for all the runs presented in Figure \ref{fig:data_scale_law} and Table \ref{table:perf_flamingo}, as detailed in Table \ref{tab:hyperparameter_details}. The training of \texttt{IDEFICS} uses a larger batch size and examples of longer sequence length. In all experimental runs, we employ the AdamW optimizer \citep{adamw} and incorporate an auxiliary loss, denoted as $ z\_loss = 10^{-3}\times log^{2}(Z) $, to encourage the softmax normalizer $log(Z)$ to get closer to 0 \citep{chowdhery2022palm}. We use gradient clipping of $1.0$.

\begin{table}[!ht]
    \centering
    \begin{tabular}{p{0.15\linewidth}p{0.3\linewidth}p{0.18\linewidth}p{0.18\linewidth}}
    \toprule
        \rule{0pt}{10pt} ~ & \textit{Parameters} & \texttt{IDEFICS-80B} & \rule{0pt}{10pt}\texttt{IDEFICS-9B} \rule[-5pt]{0pt}{0pt}\\ \hline
        \rule{0pt}{10pt}\textbf{Perceiver Resampler} & \textit{Number of Layers} & 6 & 6 \rule[-5pt]{0pt}{0pt}\\ \cline{2-4}
        \rule{0pt}{10pt}~ & \textit{Number of Latents} & 64 & 64 \rule[-5pt]{0pt}{0pt}\\ \cline{2-4}
        \rule{0pt}{10pt}~ & \textit{Number of Heads} & 16 & 16 \rule[-5pt]{0pt}{0pt}\\ \cline{2-4}
        \rule{0pt}{10pt}~ & \textit{Resampler Head Dimension} & 96 & 96 \rule[-5pt]{0pt}{0pt}\\ \hline
        \rule{0pt}{10pt}\textbf{Model} & \textit{Language Model Backbone} & Llama-65b & Llama-7b \rule[-5pt]{0pt}{0pt}\\ \cline{2-4}
        \rule{0pt}{10pt}~ & \textit{Vision Model Backbone} & \texttt{laion/CLIP-ViT\allowbreak -H-14-laion2B\allowbreak -s32B-b79K} & \texttt{laion/CLIP-ViT\allowbreak -H-14-laion2B\allowbreak -s32B-b79K}  \rule[-5pt]{0pt}{0pt}\\ \cline{2-4}
        \rule{0pt}{10pt}~ & \textit{Cross-Layer Interval} & 4 & 4 \rule[-5pt]{0pt}{0pt}\\ \hline
        \rule{0pt}{10pt}\textbf{Training} & \textit{Sequence Length} & 1024 & 1024 \rule[-5pt]{0pt}{0pt}\\ \cline{2-4}
        \rule{0pt}{10pt}~ & \textit{Effective Batch Size} (\# of tokens) & 3.67M & 1.31M \rule[-5pt]{0pt}{0pt}\\ \cline{2-4}
        \rule{0pt}{10pt}~ & \textit{Max Training Steps} & 200K & 200K \rule[-5pt]{0pt}{0pt}\\ \cline{2-4}
        \rule{0pt}{10pt}~ & \textit{Weight Decay} & 0.1 & 0.1 \rule[-5pt]{0pt}{0pt}\\ \cline{2-4}
        \rule{0pt}{10pt}~ & \textit{Optimize}r & Adam(0.9, 0.999) & Adam(0.9, 0.999) \rule[-5pt]{0pt}{0pt}\\ \cline{2-4}
        \rule{0pt}{10pt}~ & \textit{Gradient Clipping} & 1.0 & 1.0 \rule[-5pt]{0pt}{0pt}\\ \cline{2-4}
        \rule{0pt}{10pt}~ & \textit{Z-loss weight} & 1e-3 & 1e-3 \rule[-5pt]{0pt}{0pt}\\ \hline
        \rule{0pt}{10pt}\textbf{Learning Rate} & \textit{Initial Max} & 5e-5 & 1e-5 \rule[-5pt]{0pt}{0pt}\\ \cline{2-4}
        \rule{0pt}{10pt}~ & \textit{Initial Final} & 3e-5 & 6e-6 \rule[-5pt]{0pt}{0pt}\\ \cline{2-4}
        \rule{0pt}{10pt}~ & \textit{Decay Schedule} & Linear & Linear \rule[-5pt]{0pt}{0pt}\\ \cline{2-4}
        \rule{0pt}{10pt}~ & \textit{Linear warmup} Steps & 2K & 2K \rule[-5pt]{0pt}{0pt}\\ \hline
        \rule{0pt}{10pt}\textbf{Large-scale Optim}. & \textit{Gradient Checkpointing} & True & True \rule[-5pt]{0pt}{0pt}\\\cline{2-4}
        \rule{0pt}{10pt}~ & \textit{Precision} & Mixed-pres bf16 & Mixed-pres bf16 \rule[-5pt]{0pt}{0pt}\\ \cline{2-4}
        \rule{0pt}{10pt}~ & \textit{ZeRO Optimization} & Stage 3 & Stage 3 \\ \bottomrule
    \end{tabular}
\vspace{0.5em}
\caption{Training Hyper-Parameters}
\label{tab:hyperparameter_details}
\end{table}

During the training, two models -- \texttt{IDEFICS} and the 9B-parameter model trained on \texttt{LAION + OBELICS} -- encountered unrecoverable loss spikes. As a remedial measure, we restarted the training from a checkpoint before the spike, shuffled the data and optionally reduced the learning rate. Both models underwent exactly three restarts within the training duration.

The four runs conducted have distinct data mixtures as detailed in Table \ref{tab:data_mix_details}, and Tabel \ref{tab:data_source_details} gives the number of tokens and images in the different datasets. Each run involves training on a mixture of web documents and image-text pairs. A sampling probability \(p\) determines the mixture of these two data sources, which influences the frequency of batches originating from web documents versus those from image-text pairs.

For \texttt{IDEFICS} and \texttt{IDEFICS-9B}, the web-document dataset includes both \texttt{OBELICS} and Wikipedia, and the image-text pair dataset included \texttt{LAION} and Public Multimodal Dataset (PMD) \citep{FLAVA}. Given  Wikipedia and PMD's higher quality but lower number of examples, we repeat PMD three times and Wikipedia three times.

We used a deduplicated version of \texttt{LAION} \citep{DedupLAION} for all the runs where this dataset was used.

\begin{table}[!ht]
\centering
\begin{tabular}
{p{0.1\textwidth}|p{0.3\textwidth}|p{0.15\textwidth}|p{0.15\textwidth}|p{0.075\textwidth}}
\toprule
\textbf{Data Source} & \textbf{Data Type} & \textbf{\# Tokens in Source} & \textbf{\# Images in Source} & \textbf{Epochs} \rule[-5pt]{0pt}{0pt} \\ \hline
\rule{0pt}{10pt}\texttt{OBELICS} & Unstructured Multimodal Web Documents & 114.9B & 353M & 1 \rule[-5pt]{0pt}{0pt} \\ \hline
\rule{0pt}{10pt}Wikipedia & Unstructured Multimodal Web Documents & 3.192B & 39M & 3 \rule[-5pt]{0pt}{0pt} \\ \hline
\rule{0pt}{10pt}LAION & Image-Text Pairs & 29.9B & 1.120B & 1 \rule[-5pt]{0pt}{0pt} \\ \hline
\rule{0pt}{10pt}PMD & Image-Text Pairs & 1.6B & 70M & 3 \\
\bottomrule
\end{tabular}
\vspace{0.5em}
\caption{Number of tokens and images in the different datasets used for the training of \texttt{IDEFICS}.}
\label{tab:data_source_details}
\end{table}

\begin{table}[h]
\centering
\begin{tabular}{lccccccc}
\toprule
\textbf{Model} & \texttt{\textbf{OBELICS}} & \textbf{Wikipedia} & \texttt{\textbf{LAION}} & \textbf{PMD} \rule[-5pt]{0pt}{0pt} \\ \hline
\rule{0pt}{10pt}9B-parameter model, \texttt{OBELICS + LAION} &  50\% & 0\% & 50\% & 0\%\\
9B-parameter model, \texttt{OBELICS} only &  100\% & 0\% & 0\% & 0\%\\
9B-parameter model, \texttt{LAION} only &  0\% & 0\% & 100\% & 0\%\\
\texttt{IDEFICS-9B} &  73.85\% & 6.15\% & 17.18\% & 2.82\%\\
\texttt{IDEFICS} &  73.85\% & 6.15\% & 17.18\% & 2.82\%\\
\bottomrule
\end{tabular}
\vspace{0.5em}
\caption{Breakdown of the dataset mixtures used. Percentages correspond to the effective number of tokens seen from each dataset.}
\label{tab:data_mix_details}
\end{table}

\subsubsection{Compute Details}
We train the 9B-parameter models on \texttt{OBELICS}-only and \texttt{LAION}-only on 32 80GB A100 GPUs, and on \texttt{OBELICS + LAION} on 64 80GB A100s, for approximately 6 days. These 3 trainings have the same effective batch size. We train \texttt{IDEFICS} on 512 80GB A100 GPUs and \texttt{IDEFICS-9B} on 128 80GB A100 GPUs  for about 14 days each. The compute infrastructure is hosted on an AWS cluster located in Oregon.

\subsubsection{Evaluation}
To ensure fair comparisons against Flamingo \citep{Flamingo}, we make sure that we are using the same evaluation splits for each benchmark. We evaluate the models using an in-context learning approach \citep{GPT3}, with random in-context examples. For the 0-shot evaluations, as in \citet{Flamingo}, we use 2 random priming in-context examples but without passing the associated images. We systematically use different data splits to select the best-performing prompt (which involves creating validation sets from the training sets, following the methodology proposed by \citet{Flamingo}). Table \ref{tab:evaluation_prompts} lists the prompts used for each model and task.

For the classification tasks (HatefulMeme \citep{hatefulmeme}, IIIT-5k \citep{iiit5k}), we use rank classification, i.e. we compute the log probability of the prompt followed by each of the labels individually, and select as the predicted label the one with the highest probability.

For the image captioning (COCO \citep{coco}, Flickr30k \citep{flickr30k}) and visual question answering tasks (VQAv2 \citep{VQA}, OKVQA \citep{okvqa}, TextVQA \citep{textvqa}, VizWiz \citep{VizWiz}), we report evaluation in the open-ended setup. We use the greedy decoding as we found that it increased the performance. However, we observe that the models tend to generate long answers. To truncate the generated caption or answer, unless specified otherwise, we use a list of manually selected stop words. For VisDial, since the evaluation metric is NDCG, we instead rank the possible candidates for each question.

The VQA tasks comporting a high proportion of questions with a single-word answer, it was beneficial for the 9B-parameter model trained on \texttt{LAION} only to keep the first word of the generated answer as the prediction to boost its performance.

\begin{table}[htbp]
\centering
\tiny
\begin{tabular}{m{.08\textwidth}m{.15\textwidth}m{.25\textwidth}m{.25\textwidth}m{.14\textwidth}} 
\toprule
\textbf{Task} & \textbf{Model} & \textbf{Prefix prompt} & \textbf{Example prompt} &  \textbf{Stop words}\rule[-5pt]{0pt}{0pt} \\ \hline
\rule{0pt}{10pt}VQAv2\newline OKVQA\newline TextVQA & \texttt{IDEFICS}\newline \texttt{IDEFICS-9B}\newline9B \texttt{LAION} only\newline9B \texttt{OBELICS} only\newline9B \texttt{LAION} + \texttt{OBELICS} & \{bos\_token\}Instruction: provide an answer to the question. Use the image to answer.\textbackslash{}n & Image:\{token\_\allowbreak around\_\allowbreak image\}\{image\_\allowbreak token\}\{token\_\allowbreak around\_\allowbreak image\}Question: \{question\} Answer: \{answer\}\textbackslash{}n & "Question", "User", "Image", "task", "What", "Who", "When", "Where", "Why", "How"\rule[-5pt]{0pt}{0pt}\\ \hline
\rule{0pt}{10pt}COCO\newline Flickr30k & \texttt{IDEFICS}\newline \texttt{IDEFICS-9B}\newline9B \texttt{OBELICS} only\newline9B \texttt{LAION} + \texttt{OBELICS} & \{bos\_token\} & Image:\{token\_\allowbreak around\_\allowbreak image\}\{image\_\allowbreak token\}\{token\_\allowbreak around\_\allowbreak image\}Caption: \{caption\}\textbackslash{}n & "Caption", "Description", "User", "Image", "task"\rule[-5pt]{0pt}{0pt}\\\hline
\rule{0pt}{10pt}COCO\newline Flickr30k & 9B \texttt{LAION} only & \{bos\_token\}Instruction: provide a short caption of the input image.\textbackslash{}n & Image:\{token\_\allowbreak around\_\allowbreak image\}\{image\_\allowbreak token\}\{token\_\allowbreak around\_\allowbreak image\}Image description: \{caption\}\textbackslash{}n & "Caption", "Description", "User", "Image", "task"\rule[-5pt]{0pt}{0pt}\\\hline
\rule{0pt}{10pt}Hateful-Memes & \texttt{IDEFICS}\newline \texttt{IDEFICS-9B}\newline9B \texttt{LAION} only\newline9B \texttt{OBELICS} only\newline9B \texttt{LAION} + \texttt{OBELICS} & It's a conversation between a human, the user, and an intelligent visual AI, Bot. The user sends memes with text written on them, and Bot has to say whether the meme is hateful or not. & \{token\_\allowbreak around\_\allowbreak image\}\{image\_\allowbreak token\}\{token\_\allowbreak around\_\allowbreak image\}is an image with written "\{context\}" on it. Is it hateful? Answer: \{class\_name\}&\xmark \rule[-5pt]{0pt}{0pt} \\ \hline
\rule{0pt}{10pt}IIIT5k & 9B \texttt{LAION} only\newline9B \texttt{OBELICS} only\newline9B \texttt{LAION} + \texttt{OBELICS} &\xmark& \{token\_\allowbreak around\_\allowbreak image\}\{image\_\allowbreak token\}\{token\_\allowbreak around\_\allowbreak image\}"\{class\_\allowbreak name\}" is written on the picture.&\xmark \\ \hline
\rule{0pt}{10pt}VizWiz & \texttt{IDEFICS}\newline \texttt{IDEFICS-9B} & \{bos\_token\}Task: Answer the questions based on the image when possible, otherwise say unanswerable.\textbackslash{}n & Image:\{token\_\allowbreak around\_\allowbreak image\}\{image\_\allowbreak token\}\{token\_\allowbreak around\_\allowbreak image\}Question: \{question\} Answer: \{answer\}\textbackslash{}n& "Question", "User", "Image", "task", "What", "Who", "When", "Where", "Why", "How"\rule[-5pt]{0pt}{0pt} \\ \hline
\rule{0pt}{10pt}VisDial & \texttt{IDEFICS}\newline \texttt{IDEFICS-9B} &\xmark & \{token\_\allowbreak around\_\allowbreak image\}\{image\_\allowbreak token\}\{token\_\allowbreak around\_\allowbreak image\}\{caption\}. \{context\}\{class\_name\}.&\xmark \\ \bottomrule
\end{tabular}
\vspace{0.5em}
\caption{\rule{0pt}{12pt}We select the prompts from a pool of candidates by evaluating 5 intermediate checkpoints on the query and support validation task sets. To form the prompt with $N$ priming examples, we concatenate the prefix prompt, followed by $N$ example prompts filled with data from the priming examples, and finally the example prompt filled with data from the example to be evaluated. The data to be replaced is between curly brackets.}
\label{tab:evaluation_prompts}
\end{table}


\subsubsection{Additional Experimental Results}\label{sec:add_exp_results}

\input{plots/plot_detail_data_laws}

In Figure \ref{fig:detail_data_scale_law}, we plot the performance per benchmark for the 9B-parameter models trained on \texttt{LAION} only, \texttt{OBELICS} only, and a mixture of \texttt{OBELICS} and \texttt{LAION}. We notice that, even if the training on \texttt{LAION} only is smooth and the loss keeps decreasing (there are no spikes nor instabilities), performance starts to decrease after a certain point on visual question answering benchmarks. We hypothesize that training on image-text pairs can allow a fast association of concepts between images and texts, but fails to teach the model more complex reasoning skills required to solve visual question answering. We tried many different prompt candidates in order to boost the performance of the model trained on \texttt{LAION} only for the VQA tasks, without much success.

On the other hand, we note that training on image-text pairs yield stronger performance on image captioning tasks than on multimodal documents only. This is expected since training and evaluation correspond to the exact same task.

\clearpage

\newpage

\subsection{License and Author Statement}

We release the dataset under a CC-BY license and Terms of Use that require disclosure of when the dataset is used for the purpose of training models. This license is not intended to replace the licenses of the source content, and any use of content included in the dataset must comply with the original licenses and applicable rights of its data subjects.

The purpose of this statement is to clarify the responsibilities and liabilities associated with the use of this dataset. While we have made every effort to ensure the accuracy and legality of the data contained within this dataset, we cannot guarantee its absolute completeness or correctness.

Therefore, if any rights, legal or otherwise, are violated through this dataset, including but not limited to copyright infringement, privacy violations, or misuse of sensitive information, we, the authors, assume no liability for such violations.

By utilizing this dataset, you agree that any consequences, legal or otherwise, arising from using this dataset will be the user's sole responsibility. You acknowledge that you will exercise due diligence and adhere to all applicable laws, regulations, and ethical guidelines when using the dataset.

By accessing, downloading, or using this dataset, you signify your acceptance of this statement and your commitment to abide by the terms and conditions of the CC-BY license.

If you disagree with the terms of this statement or the CC-BY license, you are not authorized to use this dataset.

The dataset will be hosted and maintained on the Hugging Face Hub.

\end{document}